\documentclass[12pt]{iopart}

\usepackage{graphicx}
\begin{document}

\title{Equation of state of strongly interacting matter: spectra
for thermal particles and intensity
correlation of thermal photons}
\title[Equation of state of strongly interacting matter...]{}

\author{Somnath De\footnote{somnathde@veccal.ernet.in}, 
Dinesh K. Srivastava\footnote{dinesh@veccal.ernet.in}, 
and Rupa Chatterjee\footnote{rupa@veccal.ernet.in, Present address: 
Department of Physics, University of Jyv\"askyl\a", Jyv\"askyl\"a, Finland}}

\address{Variable Energy Cyclotron 
Centre, 1/AF Bidhan Nagar, Kolkata 700 064, India}
\begin{abstract}
We find that an equation of state for hot hadronic matter consisting of
all baryons having $M < 2$ GeV and all mesons having $M < 1.5$ GeV, 
along with Hagedorn resonances in thermal and chemical equilibrium, matches
rather smoothly with lattice equation of state (p4 action, ${N_\tau}=8$)
for T up to $\approx 200$ MeV, when corrected for the finite volume 
of hadrons.
 
Next we construct two equations of state for strongly interacting matter; 
one, HHL, in which the above is matched to the lattice equation of state
at $T=165$ MeV and the other, HHB, where we match it to a bag model
equation of state with critical temperature $T_c=165$ MeV.
We compare particle spectra, thermal photon spectra and histories of
evolution of the quark-gluon plasma produced in the central collision of 
gold (lead) nuclei at RHIC (LHC) energies, considering ideal hydrodynamical
expansion of the system.
The particle and thermal photon spectra are seen to differ only
marginally, for the two equations of state. The history of evolution shows
differences in the evolution of temperature and radial velocity, as one
might expect.

We calculate intensity interferometry of thermal photons and find it
to be quite distinct for the two equations of state, especially
for the outward correlation. The longitudinal correlation
also shows a dependence on the equation of state, though, 
to a smaller extent.
\end{abstract}

\maketitle

\section{Introduction}
The primary goal of colliding heavy nuclei at relativistic energies 
is to study the behaviour of quantum chromodynamics (QCD) at high 
energy density.
Relentless theoretical efforts over more than two decades have
given rise to a description where the two Lorentz contracted nuclei pass 
through each other and multiple parton-parton scatterings and 
a vehement production of partons take place.
This may lead to a deconfined state of quarks
and gluons in local thermal equilibrium, called quark-gluon plasma (QGP).
The unique importance of these studies lies in the well accepted
premise that the early universe was in the form of QGP at around a few 
microseconds after the Big Bang. Recent experiments at the Relativistic
Heavy Ion Collider (RHIC) at Brookhaven National Laboratory have
provided a clear proof of the formation of QGP~\cite{white}
 or rather a strongly
interacting QGP (sQGP), which behaves almost like an ideal fluid~\cite{roy}.
Notable confirmations include observations of jet quenching~\cite{jet1,jet2},
 elliptic flow~\cite{fl1,fl2}, recombination of partons as a process
of hadronization~\cite{rec},
and electromagnetic radiations~\cite{phot}. 
The theoretical descriptions include 
parton-cascade model, which represents the multiple collisions and 
multiplication of the
cascading  partons using a transport theory~\cite{pcm}.
 This has been supplemented with a hadronic cascade~\cite{bass_dum},
once the density of partons is sufficiently small. The initial state has
also been modelled as a colour gluon condensate~\cite{cgc}.

If a quark gluon plasma is formed in such collisions 
at some initial time, then assuming a
continued thermal and chemical equilibrium, the entire evolution of
the system can be calculated using relativistic hydrodynamics~\cite{dks_hyd}. 
The condition of chemical equilibrium has been relaxed in several
studies, with interesting results~\cite{biro,sspc,munshi}.
The consequence of viscosity, in the QGP or in the hadronic phase,
is also a subject of considerable interest~\cite{visc}.

Our goal in the present study is rather modest but well defined.
We consider that relativistic collision of heavy nuclei (at RHIC and
LHC energies) leads to formation of quark gluon plasma at time
$\tau_0$. The plasma then expands, cools, and hadronizes either in
a rapid cross-over as suggested by lattice QCD calculations
or in a first order phase transition at $T=T_c$, used in  a large number 
of calculations over the last two decades using  a bag model equation
of state (EOS). 
We model the hadronic phase as consisting of 
all mesons having $M< 1.5$ GeV and all baryons having 
 $M < 2$ GeV along with or without Hagedorn resonances.

 We find that
the inclusion of Hagedorn resonances and 
the finite volume for the hadrons leads to
thermodynamic quantities which join rather smoothly to those from a
lattice equation of state (p4 action; $N_{\tau}=8$)~\cite{bazavov}
for zero baryonic chemical potential. We get two different equations
of state by switching over  from this 
hadronic state to either the lattice equation
of state (HHL) or  to a bag model equation of state (HHB) at $T=$ 
165 MeV.
The bag model EOS admits a mixed phase at $T=$ 165 MeV. 
 The lattice equation of state, on the other hand,
 displays a sharp cross-over 
for $180 < T < 190 $ MeV.

We calculate thermal particle production, thermal photon production and 
the history of evolution of the system using the two equations of state HHL
and HHB for central collision of gold (lead) nuclei
at the highest RHIC (LHC) energy.
 The particle and photon spectra are seen to be only 
marginally dependent on the EOS. 

The history of evolution shows noticeable differences for
temperature and radial velocity as one might expect. Recalling that
the intensity interferometry of thermal photons is sensitive to the
history of evolution of the system, we look at the outward, side-ward,
and longitudinal intensity correlation of thermal photons having
 $K_T \leq$ 2 GeV. Again we find only a marginal difference in 
side-ward and longitudinal correlations. 

The outward correlation of thermal photons
is seen to clearly distinguish between the two equations of state.

The paper is organized as follows. In the next section we discuss the
construction of the two equations of state HHL and HHB. In section III,
 the initial conditions and the results for the particle
and thermal photon spectra at RHIC and LHC energies are given. 
In section IV, we discuss the history of evolution of the 
system for the two equations
of state; as well as the source-function for the production of photons.
In section V, we give our results for intensity interferometry of thermal
photons. Finally we summarize our findings in section VI.

\section{Equation of State}
The resonance gas model relies upon the equivalence of
thermodynamic properties of an interacting gas of hadrons and 
a free gas of hadrons and their resonances~\cite{raju}.

  We consider a hadron resonance gas,
which consists of  all mesons with mass $< 1.5$ GeV,
 and all baryons with mass $< 2.0$ GeV and their
antiparticles.  We include Hagedorn
states~\cite{hag}  having  $M>$ 2 GeV. 
Thus mass spectrum  can be written as
\begin{equation}
\rho(m)=\rho_{\rm HG}(m) + \rho_{\rm HS}(m)~,
\end{equation}
where  
\begin{equation}
\rho_{\rm HG} =\sum_{i} \ g_{i} \ \delta(m-m_{i})~. 
\end{equation}
In the above, the sum runs over all the discrete hadronic states
 and their corresponding
degeneracies are given by $g_{i}$. All the results discussed in this
work assume zero baryonic chemical potential ($\mu_B=$ 0), which is quite 
reasonable for top RHIC and LHC energies.
 
We take the density of Hagedorn states from Ref.~\cite{noronha}: 
\begin{equation}
\rho_{\rm{HS}}(m) = A \, 
\frac{\exp({m/T_H})}{(m^2+m_{0}^2)^{5/4}}~,
\end{equation}
where,
$A=$ 0.5 GeV$^{3/2}$, $m_0=0.5$ GeV, $T_H=$ 0.196 GeV,
and $m$ varies from $M_0=$ 2 GeV to 
 $M_{\rm {Max}}= $ 12 GeV.
It is noted that the inclusion of the Hagedorn resonances 
is responsible for the rapid chemical equilibration of hadrons
produced in relativistic heavy ion collisions~\cite{noronha}. This may also
signal existence of actual hadronic states (see, e.g.,
Ref.~\cite{godbole,bm,marco}) which have not yet been clearly 
identified.
Only mesonic Hagedorn states are considered (as in the above studies) and various thermodynamic quantities like - pressure, energy density, number density, and entropy density  of the hadronic gas are calculated using standard methods of statistical mechanics. 

Several procedures~\cite{jean,prakash,raf,joe} have been used in the
literature to account for the finite volume occupied by the hadrons. The treatment advocated by Kapusta and Olive~\cite{joe} is
believed to be thermodynamically consistent and has been used for this study. 
We shall see that this correction
plays an important role in getting results in conformity with those from 
lattice calculations.
According to this treatment, the finite or excluded volume (xv)
 corrected pressure, 
temperature, energy density, and entropy density are related to those 
for the point-particle (pt) values as:
\begin{equation}
 p_{\rm xv}=\frac{p_{\rm pt}(T^*)}{1-\frac{p_{\rm pt}(T^*)}{4B}} ~,
\end{equation}

\begin{equation}
 T_{\rm xv}=\frac{T^*}{1-\frac{p_{\rm pt}(T^*)}{4B}} ~,
\end{equation}
\begin{equation}
 \varepsilon_{\rm xv}=\frac{\varepsilon_{\rm pt}(T^*)} 
{1+\frac{\varepsilon_{\rm pt}(T^*)}{4B}}~,
\end{equation}
and
\begin{equation}
 s_{\rm xv}=\frac{s_{\rm pt}(T^*)}{1+\frac{\varepsilon_{\rm pt}(T^*)}{4B}}~,
\end{equation}
where $T^*$ is the temperature for the system having point particles and 
$B$ (taken as $B^{1/4}=$ 340 MeV from Ref.~\cite{noronha}) is the bag 
constant in the MIT Bag model description of hadrons.

We discuss our results for the hadronic matter with successively increasing
richness of the description; viz., hadron gas, volume corrected hadron gas,
hadron~+~Hagedorn gas, and volume corrected hadron~+~Hagedorn gas. 

The action for the lattice calculations is given by:
\begin{equation}
\frac{\Theta^{\mu\mu}(T)}{T^4}\,\equiv\,\frac{\varepsilon-3p}{T^4}\,=\,
T \, \frac{\partial}{\partial T}(p/T^4).
\end{equation}
 The parametrized form of $\Theta^{\mu\mu}(T)/T^4$ is taken 
from Ref.~\cite{bazavov}, where simulation results are given upto 
a temperature of 539 MeV. It is arbitrarily extended to 
larger values of temperature in our calculations for LHC energy.

Thus, the pressure is obtained by integrating $\Theta^{\mu\mu}/T^5$ over the
temperature,
\begin{equation}
\frac{p(T)}{T^4}-\frac{p(T_0)}{T_0^4}=\int_{T_0}^{T}\, dT^\prime \frac{1}
{T^{\prime 5}} \Theta^{\mu\mu}(T^\prime).
\end{equation}

It is found that the pressures for all the four descriptions
of the hadronic matter discussed earlier, are nearly identical at 
a temperature of about 140 MeV. Thus, for lattice results the 
final values for the pressure are obtained by taking $T_0$ as 140 MeV, 
and adding the corresponding pressure from the hadronic matter calculations.

\begin{figure}[tbp]
\begin{center}
\includegraphics[width=7.0cm, clip=true]{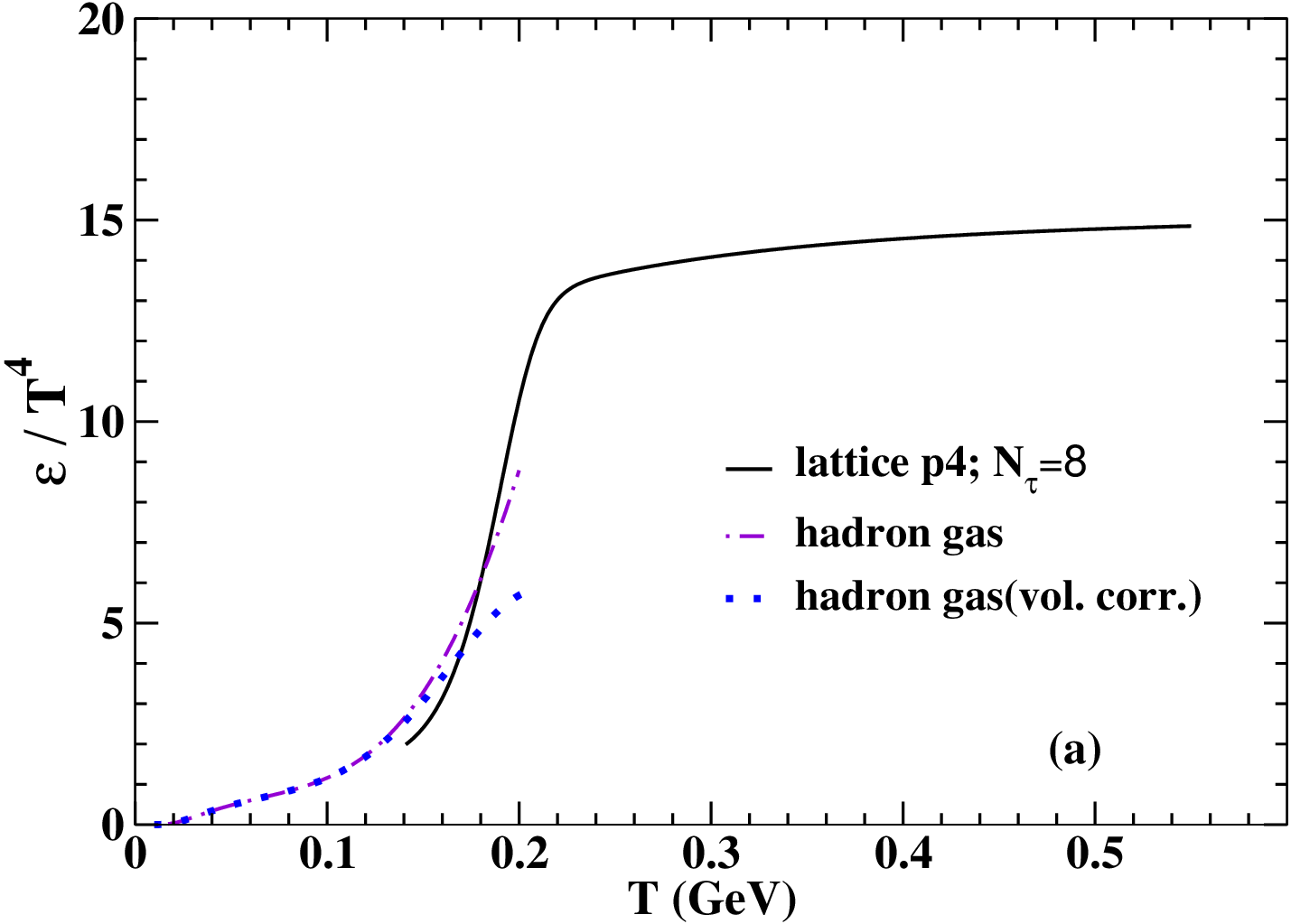}
\includegraphics[width=7.0cm, clip=true]{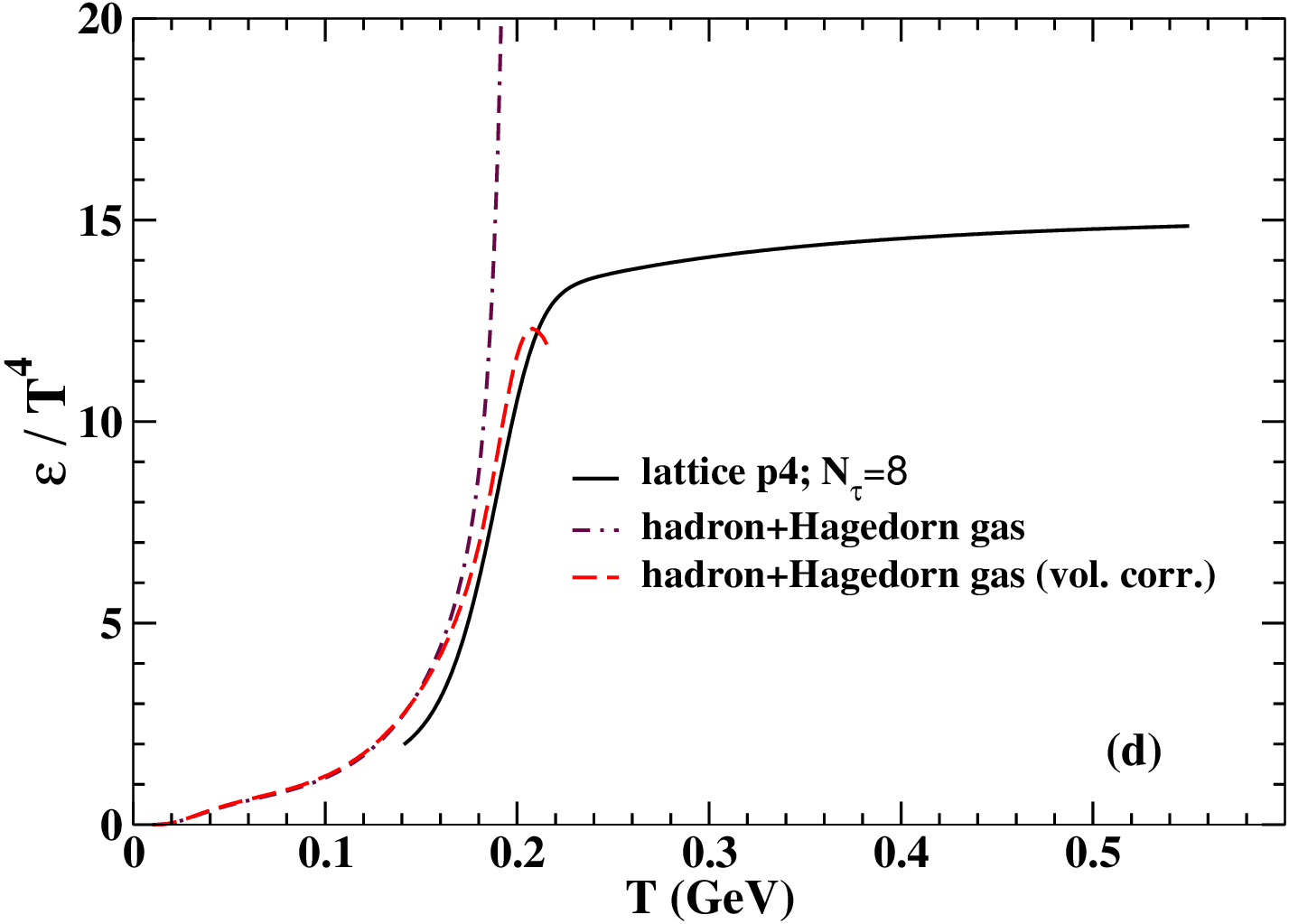}
\includegraphics[width=7.0cm, clip=true]{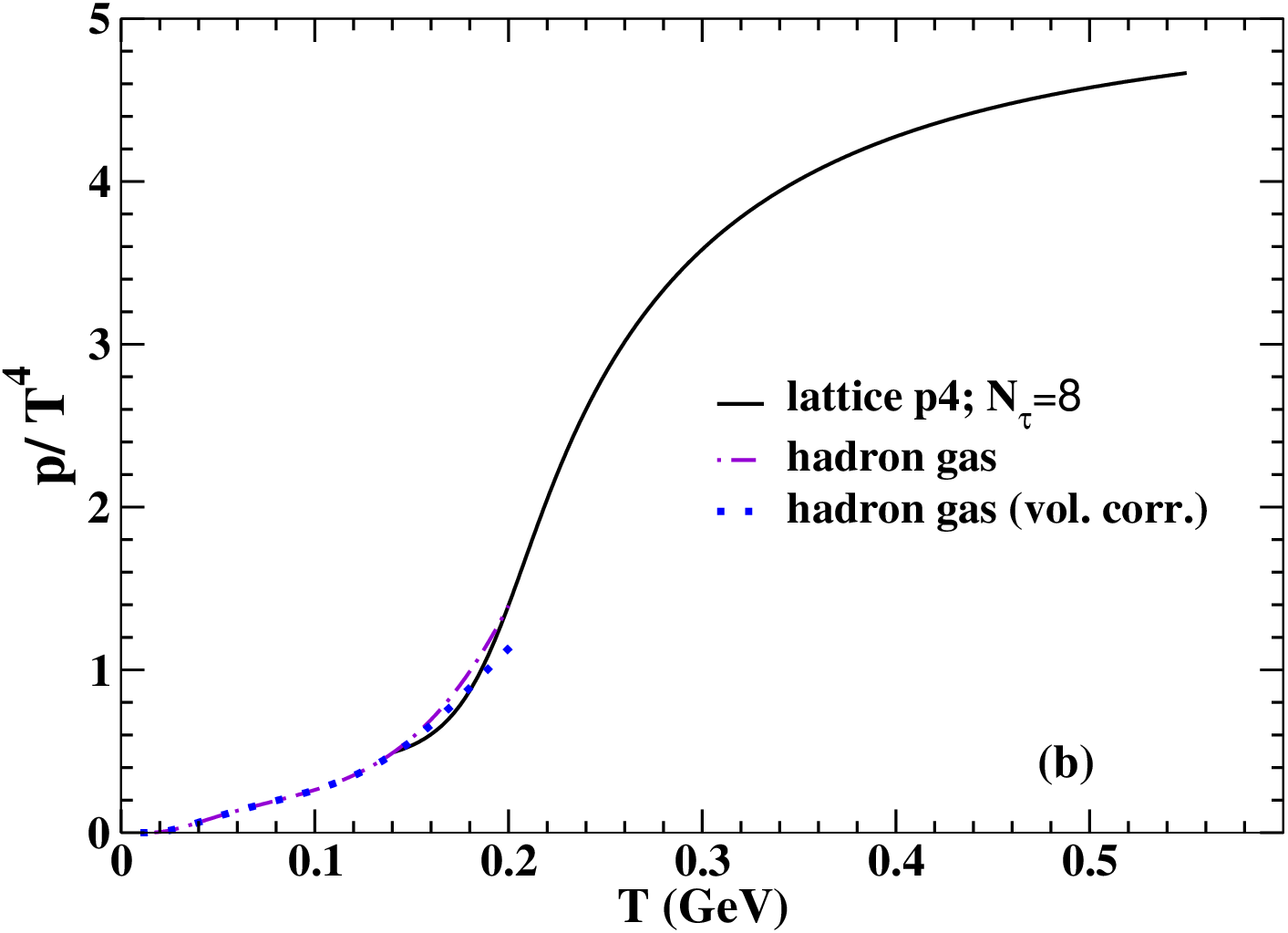}
\includegraphics[width=7.0cm, clip=true]{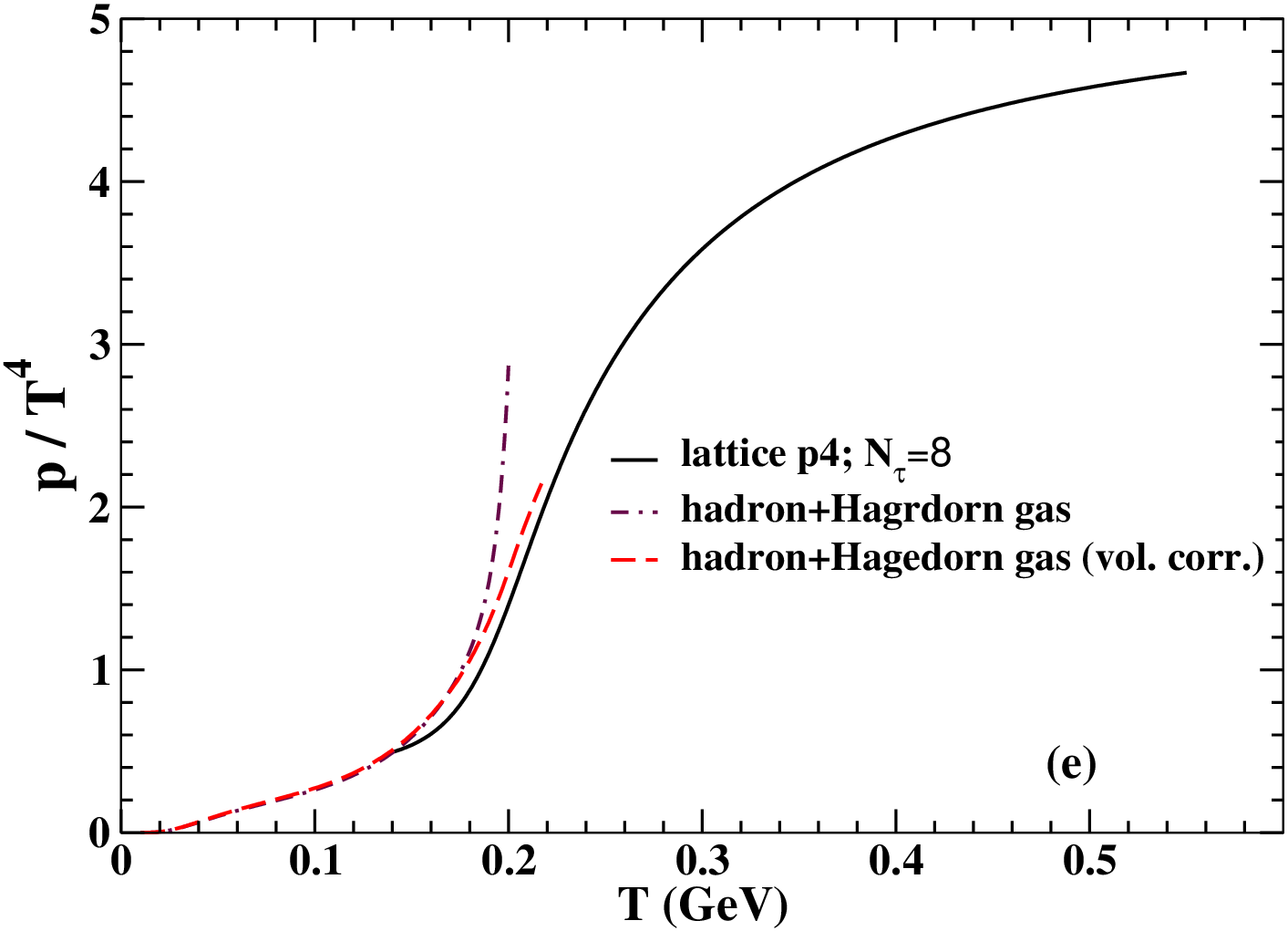}
\includegraphics[width=7.0cm, clip=true]{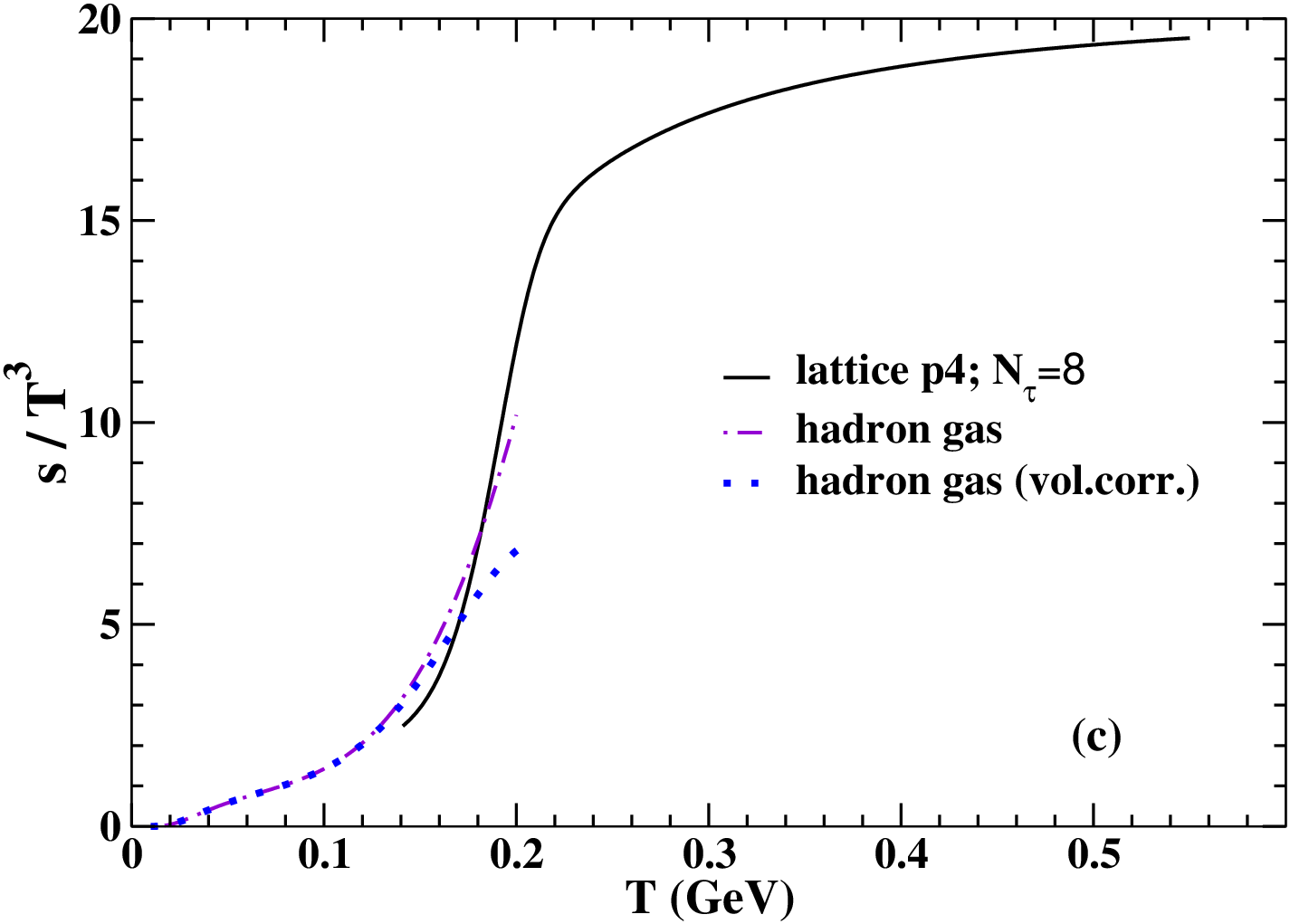}
\includegraphics[width=7.0cm, clip=true]{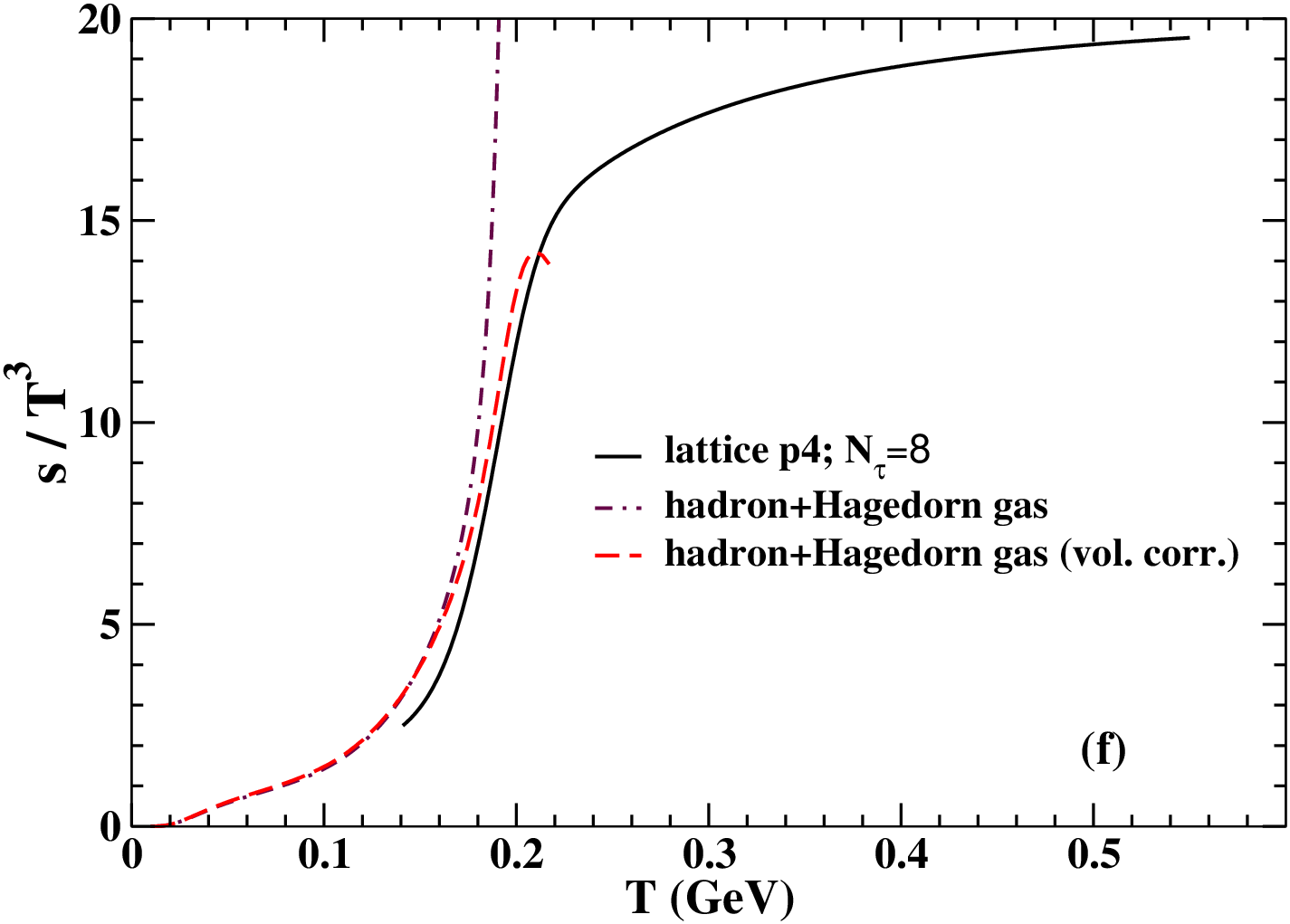}
\caption{$\varepsilon/\rm{T}^{4}$, $p/T^4$, and $s/T^3$
 for hadron gas and volume corrected hadron gas, along with lattice
results (a, b, and c). The panels d, e, and f depict corresponding
results for hadron~+~Hagedorn gas without and with correction for
finite volume of the particles.}
\label{fig1}
\end{center}
\end{figure}

In Fig.\ref{fig1} (a--f) our results for the energy density, pressure,
and entropy density are shown.

We see that the energy density, pressure, and the entropy density for the
hadronic gas (point particles) rise rapidly as the temperature 
increases beyond 160 MeV, as the number density of the
particles becomes large. Switching on the volume corrections, arrests
this rapid rise, and gives results which approach the values obtained
from lattice calculations. 

Inclusion of Hagedorn resonances gives results which are similar
in shape to those from lattice calculations over several tens of MeV.
Best agreement with the lattice calculations are obtained when the
hadron~+~Hagedorn gas is corrected for the finite volume of
the particles. Now we see a very close agreement of all the thermodynamic
quantities for the hadronic matter with those obtained using the
lattice calculations. 

In view of the above, in the following we shall describe the hadronic
matter as volume corrected hadron~+~Hagedorn gas.
  
\begin{figure}[tbp]
\begin{center}
\includegraphics[ width=9.5cm, clip=true]{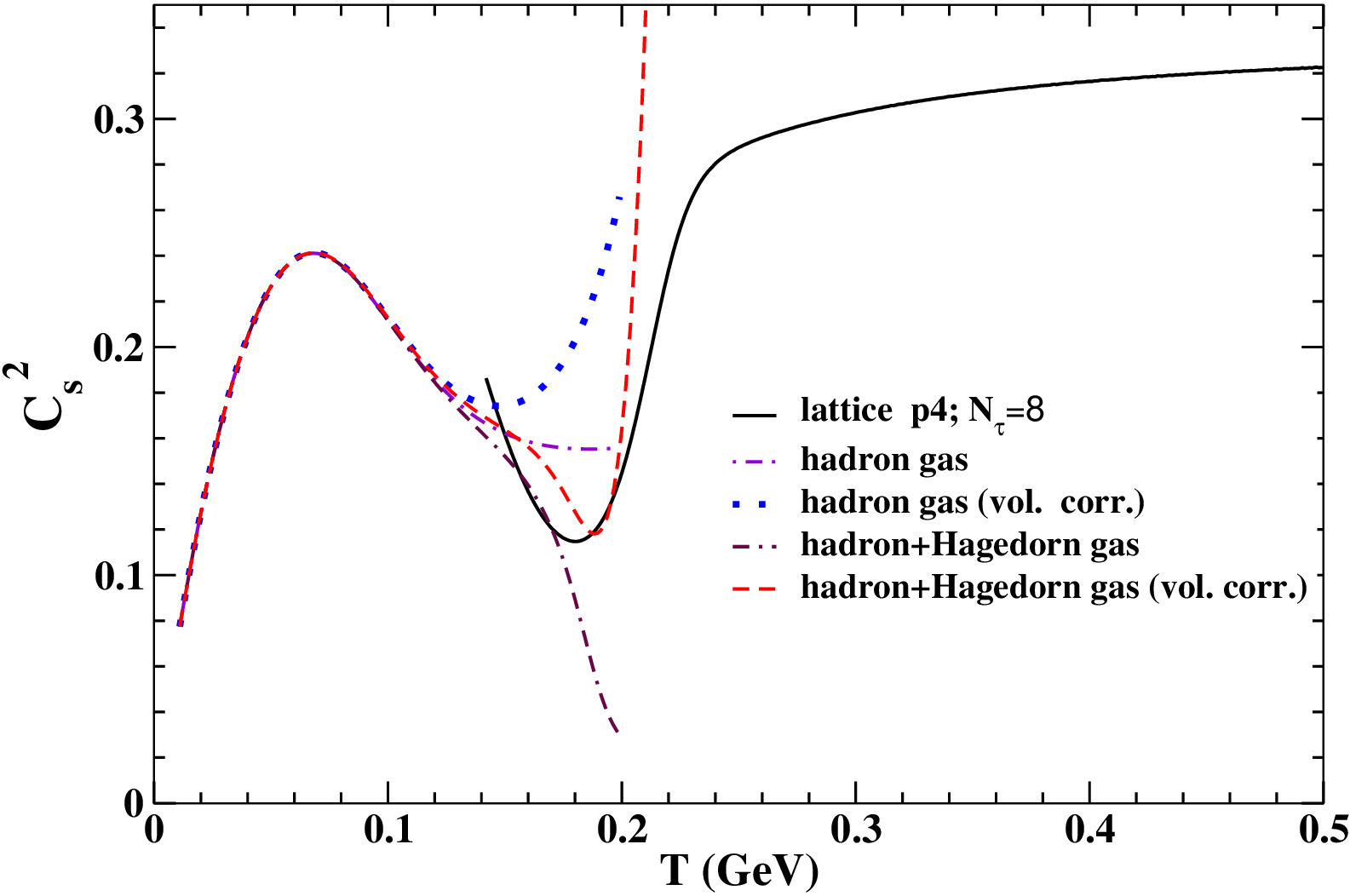}
\end{center}
\caption{(Colour on-line) Speed of sound for (i) hadron resonance gas, 
(ii) volume corrected hadron resonance gas, (iii) hadron and Hagedorn resonance gas and (iv) volume corrected hadron and Hagedorn resonance gas with
lattice results.}
\label{fig2}
\end{figure}
\begin{figure}[tbp]
\begin{center}
\includegraphics[ width=8.5cm, clip=true]{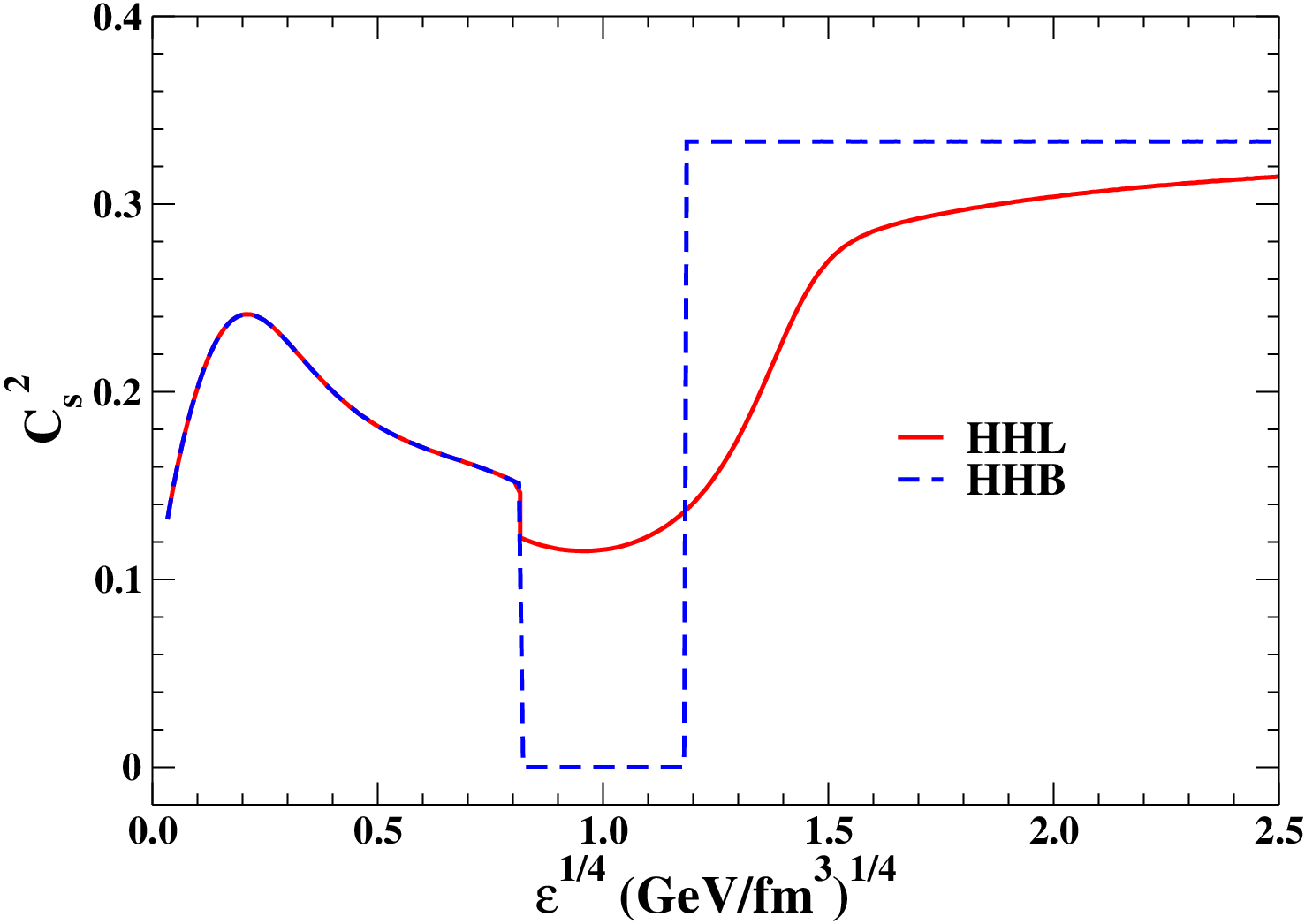}
\end{center}
\caption{(Colour on-line) Speed of sound for the 
two equations of state; HHB and HHL.}
\label{fig3}
\end{figure}

The results for the square of speed of sound for the four
descriptions of the hadronic matter and  their comparison with
the one obtained from the lattice calculations are shown in 
Fig.~\ref{fig2}. Once again 
we see that the hadron~+~Hagedorn gas with volume correction gives a speed
of sound which is close to that for the lattice calculations over a large
window of temperature. 

Now we construct two equations of state, HHL and HHB. 
For both, we use the volume corrected hadron~+~Hagedorn gas for $T<$ 165
MeV. This is motivated by the fact that the particle ratios using thermal
models at RHIC
suggest a chemical freeze-out temperature which is close to 165 MeV
at the top RHIC energies~\cite{johanna}. Similar results are expected
for the top LHC energy also~\cite{cleymans}. 

For the HHL equation of state, we switch over to the lattice equation
of state at $T=$ 165 MeV. We have found that while the variation of
 energy density, pressure, and entropy density is rather continuous 
around the point of switch over~Fig.\ref{fig1}, 
a slight discontinuity (of about 3\%)
is seen in the speed of sound. We do not expect it to cause
any noticeable difficulty in the final results.

For the HHB, we match our hadronic description to a bag model
equation of state, by adjusting the bag pressure to give the transition
temperature as 165 MeV. The matter at $T>T_c$, then consists
of non-interacting and mass-less quarks (u, d, and s) and gluons, at 
zero baryonic chemical potential. 

In Fig.~\ref{fig3} we have shown the square of the speed of sound for
the two equations of state as a function of $\varepsilon^{1/4}$, which clearly
depicts the vanishing speed of sound for the HHB equation of state,
over the energy density covered by the mixed phase. 

We reiterate that during the mixed phase,  $0.4 \leq \varepsilon \leq 2 $
 GeV/fm$^3$,
the speed of sound is zero for the HHB and is larger than that for the
HHL equation of state for higher energy densities. The speed of sound for the
HHL is never zero. These differences in the speed of
sound should lead to interesting differences in the history of
 evolution of the system, though the integrated effects may still turn out to
be small.

\section{Initial Conditions and Particle Spectra}
For 200A GeV Au+Au collision at RHIC we assume QGP is formed 
at $\tau_0=$ 0.2 fm/$c$ with an energy density used in earlier 
works~\cite{rupa} which provides  a good description to particle 
spectra and thermal photon spectra obtained experimentally. 
This is obtained by assuming that a part of
the entropy is generated in hard collisions and the remaining part is
produced in soft collisions~\cite{heinz} and the relative magnitudes 
are then adjusted to give the rapidity density of the particles \
produced in the collisions $dN_{\rm ch}/dy\approx $ 680. 
At LHC (Pb+Pb @5.5A TeV), a smaller value of $\tau_0=$ 0.1 fm/$c$ is considered and the corresponding rapidity density $dN_{\rm ch}/dy \approx $ 2040~\cite{rupa}.

For both RHIC and LHC, small value of $\tau_0$ is considered as 
we are also interested in thermal photon spectra,
which derive a large contribution from early times. If higher formation
times are assumed, then the entropy density (and the energy density) can be
correspondingly reduced.

The average initial energy densities are:

\begin{equation}
\langle \varepsilon_0 \rangle =  \left\{ 
\begin{array}{l l}
   80.8 {\rm \ GeV/fm^3}& \, {\rm at \ RHIC, \ \tau_0  \, = \, 0.2 \, fm/c}, \\
\\   
   718.3 {\rm \ GeV/fm^3}& \, {\rm at \ LHC, \ \ \, \tau_0  \, = \, 0.1 \, fm/c}. \\
\end{array} \right. \nonumber
\end{equation} 


We numerically solve the hydrodynamics equations~\cite{dks_hyd} for central collisions for a azimuthally symmetric, and longitudinally boost-invariant expansion
of the system for identical initial conditions and the two equations of
state. The freeze-out is assume to take place at $T=$ 100 MeV, and the
particle spectra are obtained using the Cooper-Frye formula~\cite{cooper}.
The photon production is calculated over the history of evolution of
the system~\cite{ss}.
The rate of production of photons from the quark matter (QM)
is taken from the complete leading 
order calculation of Arnold {\it et al.}~\cite{guy}.
The production of photons from the hadronic matter (HM) is estimated using the
results from Turbide {\it et al.}~\cite{simon}.

For the equation of state incorporating
the bag model, the quark fraction during the mixed phase
is obtained using a procedure developed earlier~\cite{kkmm}.
Let us assume that the energy density of the hadronic phase at $T=T_c$ is
$\varepsilon_h(T_c)$ and that for the quark-gluon plasma phase is
$\varepsilon_q(T_c)$. One can  easily show that~\cite{kkmm} 
a fraction $f$ of the energy density ($\epsilon(f,T_c)$)
 in the mixed phase
will be contributed by the quark-gluon plasma, where
\begin{equation}
\varepsilon(f,T_c)=\varepsilon_q(T_c)f+\varepsilon_h(T_c)(1-f)~.
\label{f}
\end{equation}
The remaining fraction $(1-f)$ will be contributed by the hadronic matter,
as seen above.
 
For the 
equation of state incorporating lattice results, we assume that the
matter consists of quarks and gluons for $T\geq$ 185 MeV and below that
it is hadronic.

\begin{figure}[tb]
\begin{center}
\includegraphics[width=7.50cm, clip=true]{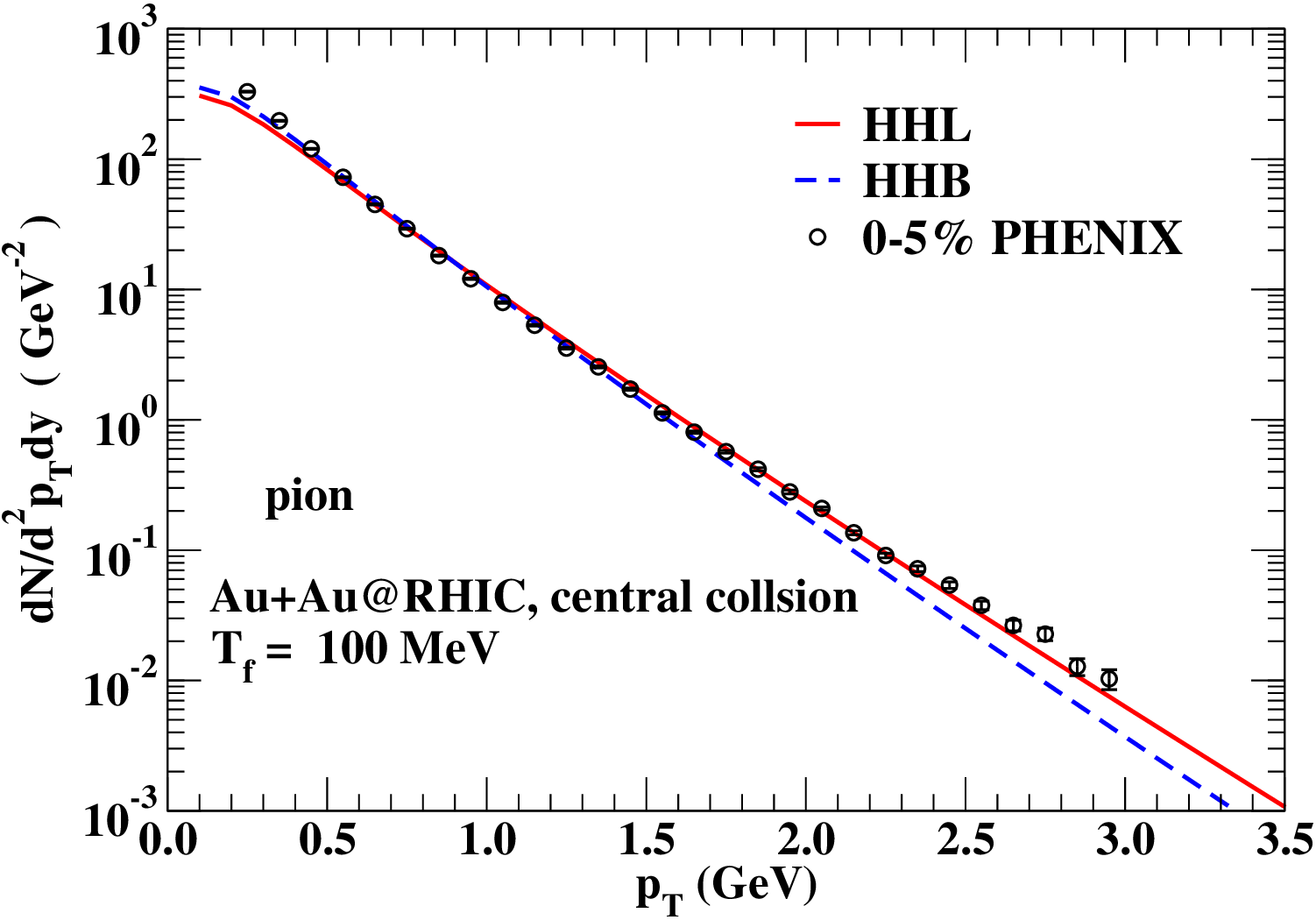}
\includegraphics[width=7.50cm, clip=true]{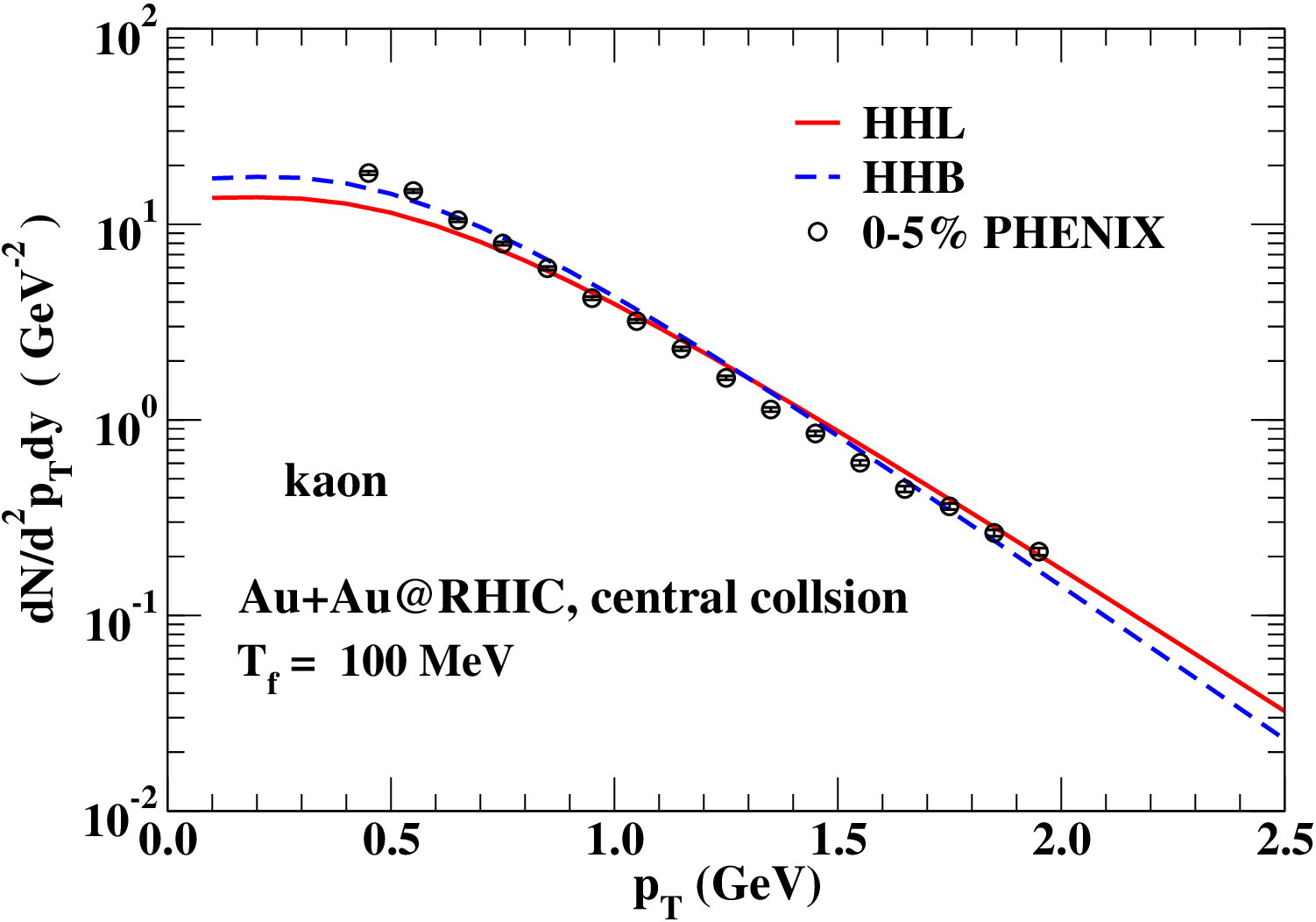}
\includegraphics[width=7.50cm, clip=true]{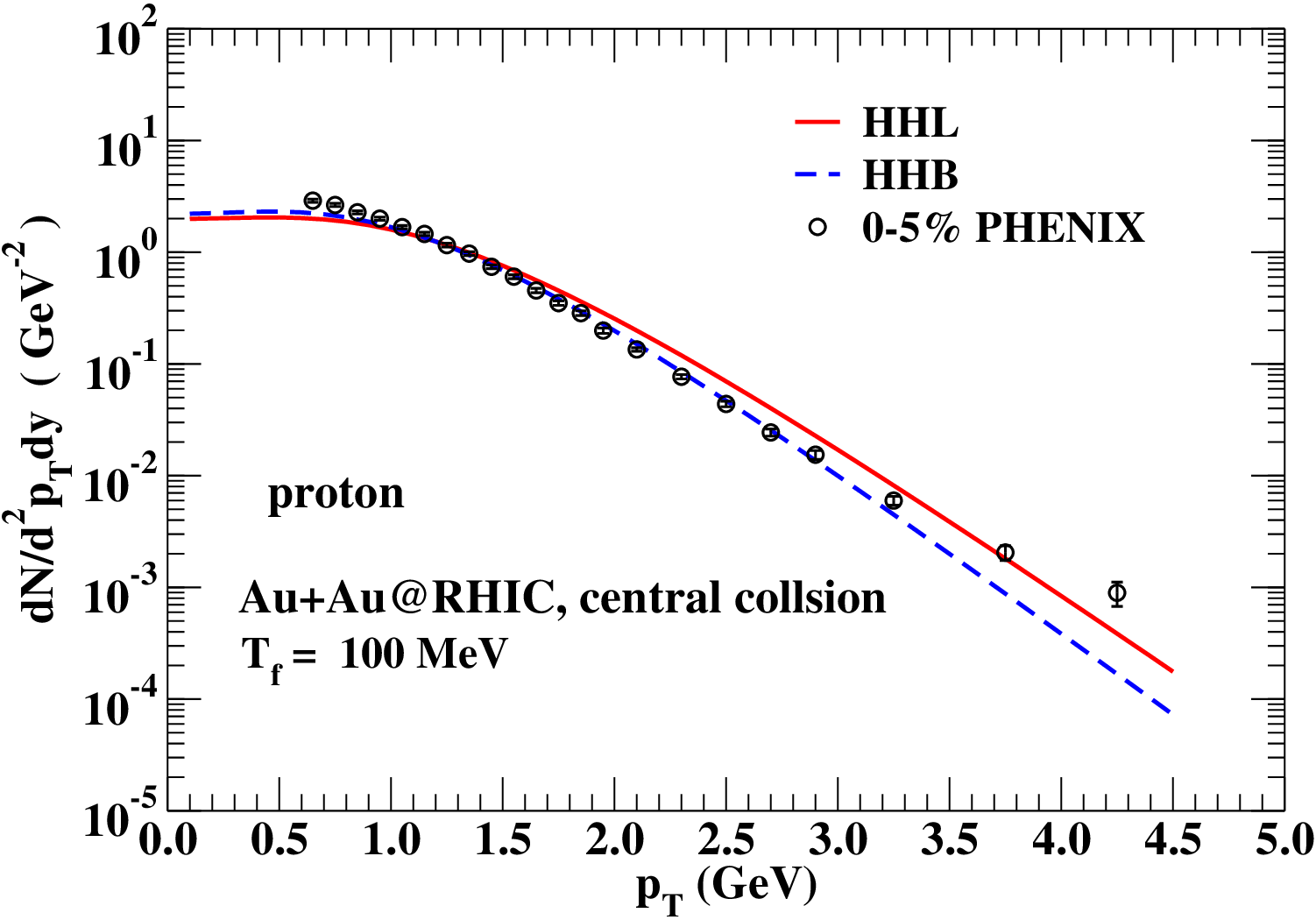}
\includegraphics[width=7.50cm, clip=true]{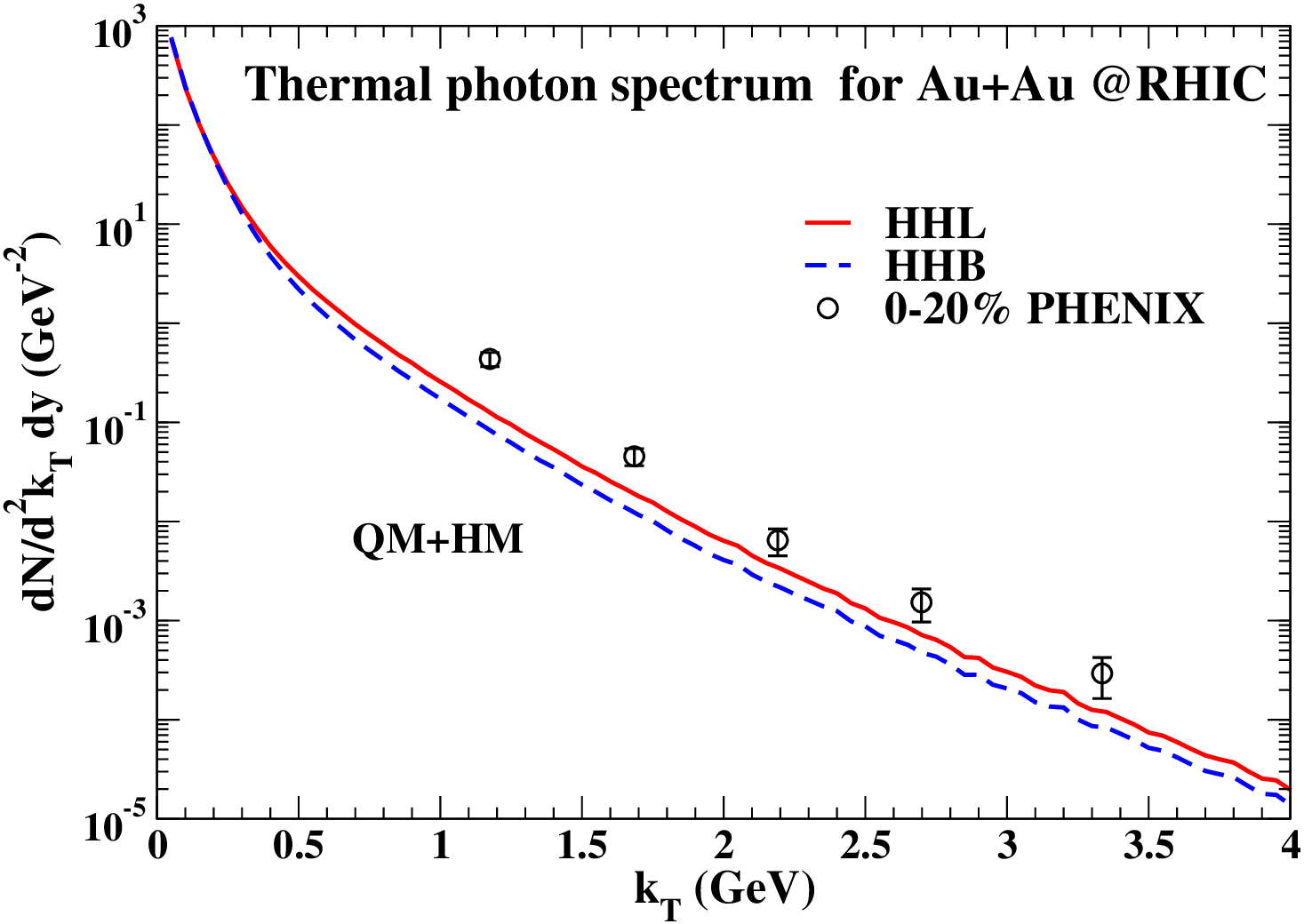}
\end{center}
\caption{(Colour on-line) 
Pion, kaon, proton and thermal photon $p_T$ spectra at RHIC for
the equations of state, HHB and HHL. All the calculations are for impact parameter b=0 fm. The experimental data for hadrons (0--5\% centrality bin) are taken from~\cite{phenix_data} and photon data for 0--20\% centrality bin from~\cite{phot_phenix} (see text for detail).}
\label{fig4}
\end{figure}

In Fig.~\ref{fig4} the results for transverse momentum distribution of thermal
photons, pions, kaons, and protons for central collision of gold
nuclei at $y=0$, for the two equations of state, at RHIC energy are shown.
 The experimental data
for 0--5\% most central collisions for the pions, kaons, and protons
(from Ref.~\cite{phenix_data}) are also shown for comparison. No attempts were made
to adjust any parameters or normalizations. 

Several interesting facts emerge.
First of all we note that both the equations of state give a reasonable
description to the particle distributions, though a slight preference
for the HHL equation of state is easily discernible. Secondly, the 
inverse slope of the spectra for HHL equation of state is larger than the same for the HHB equation of state, and
the largest difference is seen for the spectra for protons.

We feel that the main source of the difference lies in the
variation of speed of sound with energy density for the two 
equations of state. When the system follows HHB equation of 
state, the acceleration of the expansion is stalled during 
the mixed phase, which  gives rise to a smaller radial velocity 
at the time of freeze-out. 
We add that the contribution of resonance decay (which would 
improve the description of the experimental data at lower $p_T$) is not 
included in this analysis.

Similar results have been reported by authors using bag-model and
lattice equation of state~\cite{pasi} at RHIC energies. The authors
of Ref.~\cite{pasi} also report only a slight difference even for the elliptic
flow of pions, though for protons the difference is still noticeable. Their
description of the hadronic matter, however, is not as rich as that used in the
present work.

We have verified that the thermal photon spectrum is quite close to 
similar calculations reported earlier~\cite{rupa}. We have also shown the data
for single photons for 0--20\% most central collisions~\cite{phot_phenix} for a comparison. A complete description of photon data would involve addition of prompt contribution scaled by appropriate nuclear overlapping function~\cite{aurenche}.

\begin{figure}[t]
\begin{center}
\includegraphics[width=7.50cm, clip=true]{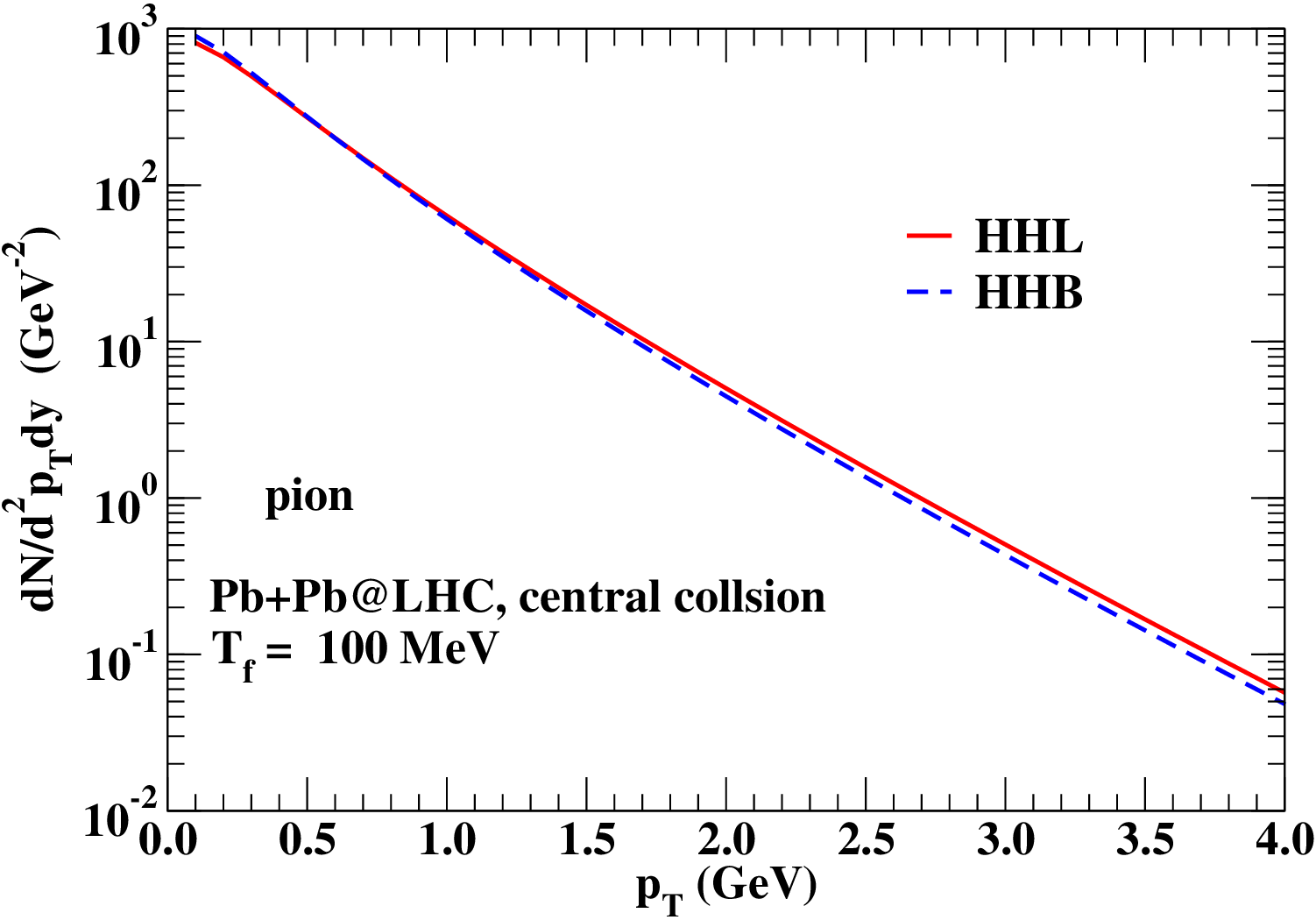}
\includegraphics[width=7.50cm, clip=true]{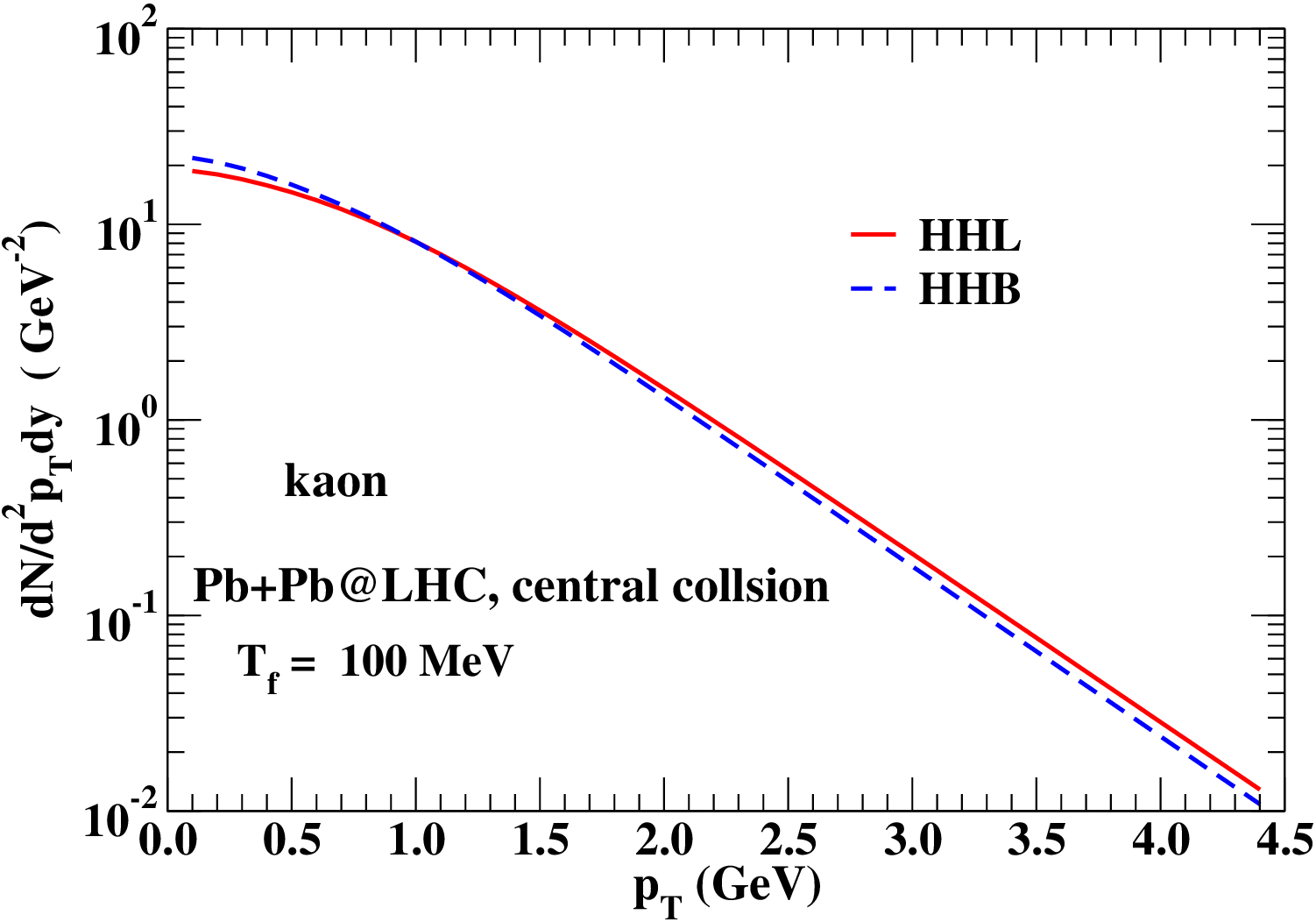}
\includegraphics[width=7.50cm, clip=true]{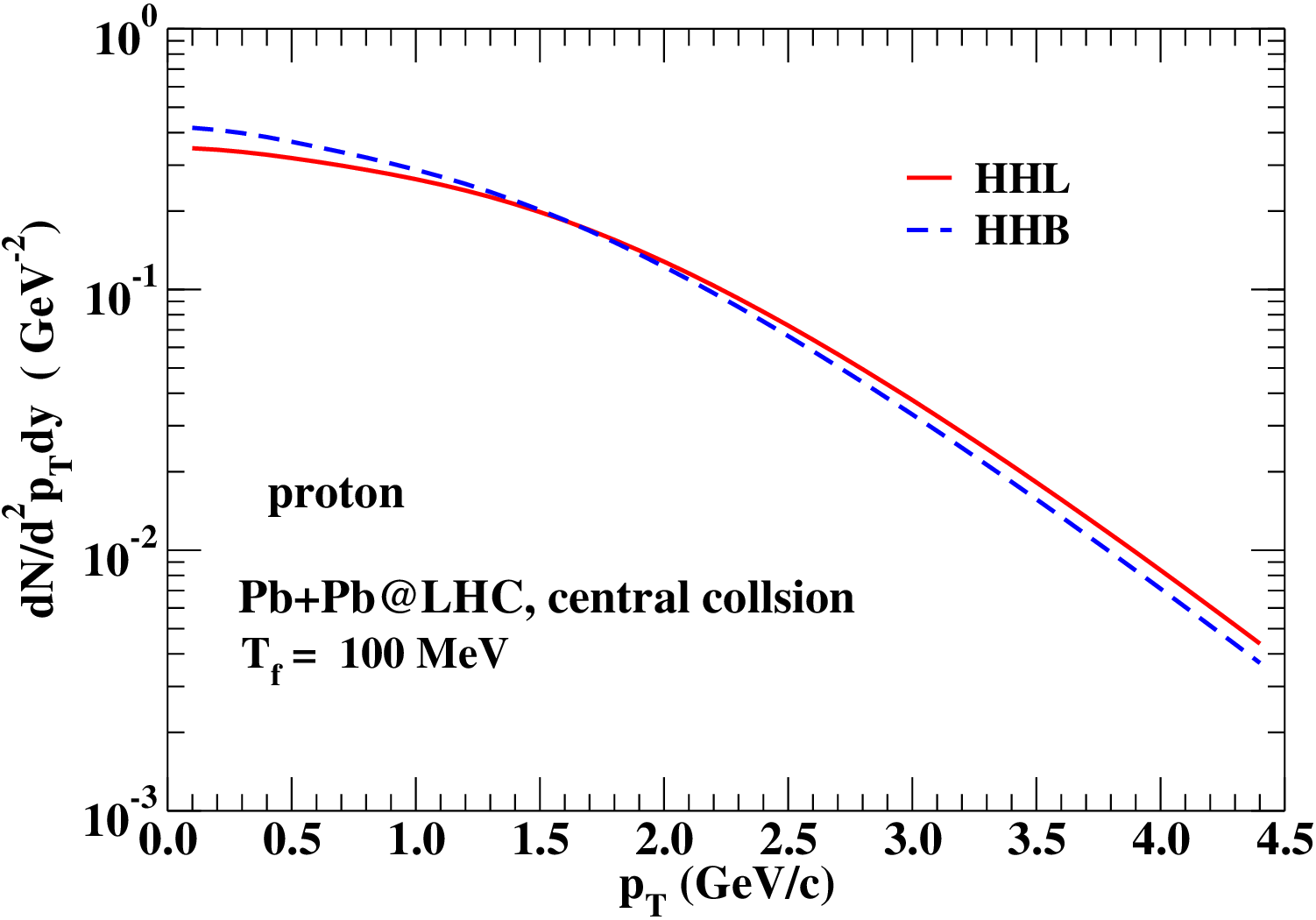}
\includegraphics[width=7.50cm, clip=true]{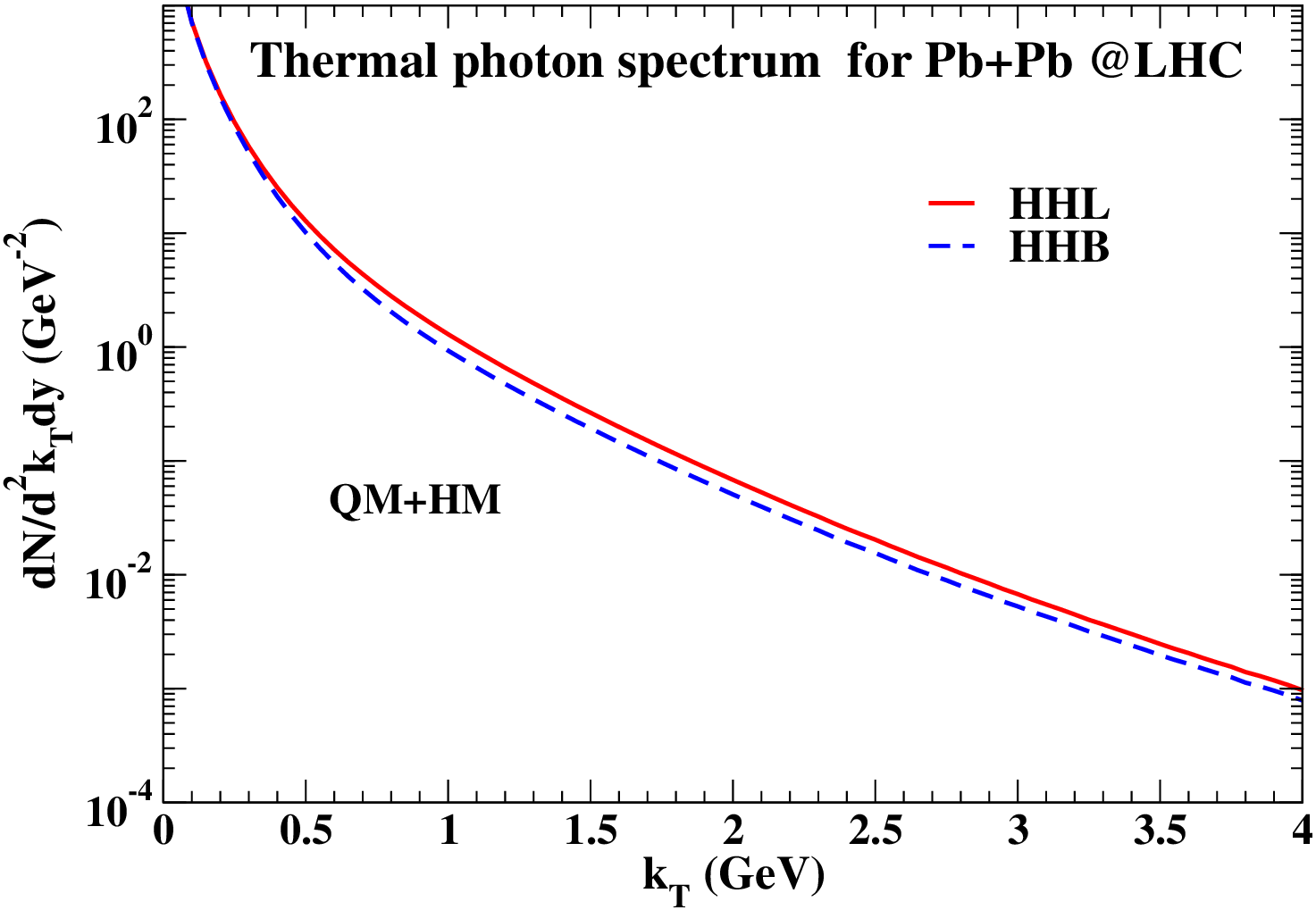}
\end{center}
\caption{(Colour on-line) Same as Fig.~\ref{fig4} at LHC.}
\label{fig5}
\end{figure}

Corresponding results for the top LHC energy are shown in Fig.~\ref{fig5}.
One can see that the difference between the transverse momentum distribution of 
photons and the hadrons for the two equations of state is
further reduced, though the slight increase in the inverse slope for the
HHL equation of state can still be seen. This suggests that the life-time of
the system is large enough to greatly obliterate the differences in
the momentum distribution of the particles at the time of freeze-out.
In fact, for both the cases the difference in the history of evolution
which should have affected the thermal photon spectra is also obliterated
due to the integration.

These results suggest that the particle spectra (even  for thermal photons) can
not distinguish between the equations of state, one admitting 
a first order phase transition, and the other admitting a rapid cross-over
as suggested by lattice calculations.

Are we to understand that there is no way to distinguish between these
two scenarios? We realise however that the history of evolution of the two systems must be different,
as they are subjected to different rates of expansion due to the
varying speed of sound.

We recall some early works~\cite{scot1,miklos} where it was suggested
that a big difference between the so-called $R_{\rm {out}}$ and
 $R_{\rm {side}}$ may imply a first order phase transition.
The experimental data at RHIC energies, however point to 
an $R_{\rm{out}}/R_{\rm{side}}\approx$ 1, which leads to  the so-called
HBT
puzzle~\cite{puzzle1,puzzle2} at RHIC. It 
has  also spawned several complex questions like what does interferometry of
pions  actually measure. A definite, though not sufficient, improvement
has been reported when a lattice inspired equation of
state is used~\cite{scot2}. It has also been suggested~\cite{scot2}
that several other small corrections add up to fully 
explain this feature.

We adopt a two-pronged approach in the following. First, we
trace the history of evolution of the systems, both at RHIC
and LHC, for the two equations of state.

Next, we study the intensity interferometry of thermal
photons, which should be free
from such complications and which should be sensitive to 
the history of the evolution
of the system. 
 We shall see that this could be quite
rewarding indeed.

\section{History of Evolution of the Collision}
\begin{figure}[tb]
\begin{center}
\includegraphics[width=7.50cm, clip=true]{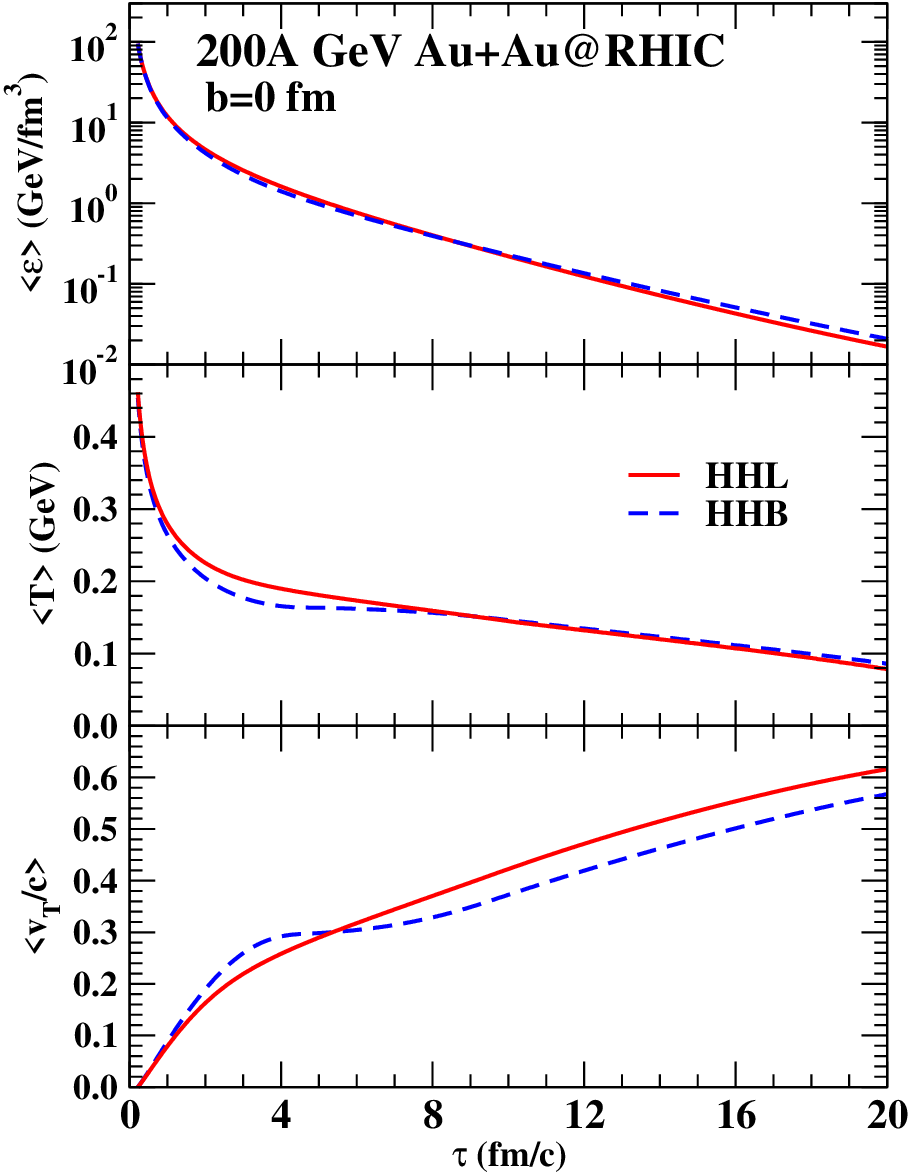}
\includegraphics[width=7.50cm, clip=true]{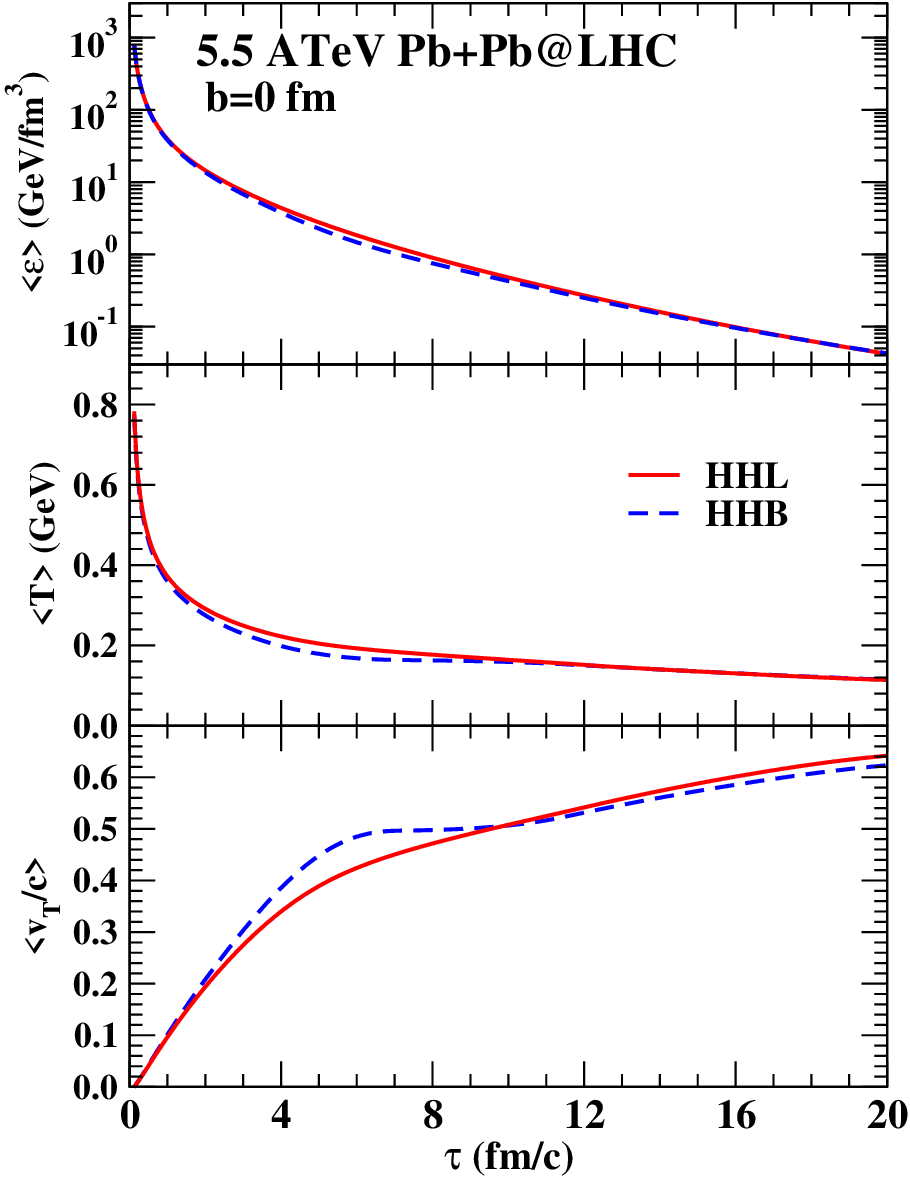}
\end{center}
\caption{(Colour on-line)
Temporal evolution of average energy density, temperature, 
and radial flow velocity at RHIC (left panel) and LHC (right panel)
for the HHB and HHL equations of state.}
\label{fig6}
\end{figure}

The energy-density weighted average temporal evolution of 
energy density, temperature, and the radial velocity
for central collisions at  RHIC and LHC energy, are shown in Fig.~\ref{fig6}
following the two equations of state,
HHB and HHL.

The average is obtained as:
\begin{equation}
\langle f \rangle =\frac{\int \, 2\pi \, r \, dr \, f(r,\tau) \,  \varepsilon(r,\tau)}
                        {\int \, 2\pi \, r \, dr \, \varepsilon (r,\tau)}~.
\end{equation}

We see (Fig.~\ref{fig6}) that both at RHIC and LHC, the variation of
the energy density with time for the two equations of state is 
quite similar, though the numerical values show a marginally rapid 
decrease in $\langle \varepsilon \rangle$ at early times for the HHB.
 This is due to a larger value of 
speed of sound in the QGP phase for the HHB equation of state
 (see. Fig.~\ref{fig3}).

The time variation of the temperature is a little more interesting as 
it stays nearly constant when bulk of the system is in the mixed phase
for the HHB equation of state, whereas it continues to decrease for the
other case.

The most interesting and potentially useful variation is observed for the
radial velocity for the two equations of state. The HHB
equation of state gives rise to a larger $v_T$ during the early 
QGP phase. The rise of the radial velocity stalls, once the system
enters a mixed phase and then rises again. The radial velocity
for the HHL equation of state rises continuously, stays below that
for the HHB equation of state, but overshoots it once the former stalls due to
the onset of the mixed phase. The final $v_T$ for the HHL EOS is
slightly larger than that for the HHB. We have already seen the
consequence of this in the particle spectra (Figs.~\ref{fig4} 
and \ref{fig5}).  

\section{Intensity Interferometry of Thermal Photons}

The intensity interferometry of thermal photons has been
suggested as a valuable probe for the history of evolution
of the system formed in relativistic heavy ion collisions~\cite{dks1,dks2,ors}.
Two recent studies (Ref.~\cite{dks3} and Ref.~\cite{uli}) have further
refined these early expectations. The authors of Ref.~\cite{dks3}
report an exciting possibility of observation of interference of
photons from the quark-phase and the hadronic phase, while Ref.~\cite{uli}
explores the details of angle-dependent intensity correlation of thermal
photons for non-central collisions.

The WA98 collaboration~\cite{wa98} has already succeeded in such a study
at SPS energy with interesting results. It is expected that the
increased initial temperature, additional sources of 
photons~\cite{bms,fms,gale},
 and jet-quenching (which suppresses the production of $\pi^0$ at larger 
$p_T$) may make this study  more feasible at RHIC and LHC energies. 

The correlation function between two photons with momenta 
$k_{1}$ and $k_{2}$
is defined in general as:
\begin{equation}
C(\mathbf{k_1},\mathbf{k_2})=  \left[ E_1E_2\frac{dN}{d^3k_1d^3k_2}\right] {/} \left[E_1\frac{dN}{d^3k_1}\,E_2\frac{dN}{d^3k_2}\right]~,
\end{equation}
where the numerator gives the correlated spectrum and the denominator
denotes the product of single spectrum of the two photons.
This can be written in terms of the source-function as
\begin{equation}
C(\mathbf{q},\mathbf{K})=1+\frac{1}{2}
\frac{\left|\int \, d^4x \, S(x,\mathbf{K})
e^{ix \cdot q}\right|^2}
                        {\int\, d^4x \,S(x,\mathbf{k_1}) \,\, \int d^4x \,
S(x,\mathbf{k_2})} ~.
\label{def}
\end{equation}
In the above, $S(x,{\bf K})$ is the space-time emission function, and 
\begin{equation}
\mathbf{q}=\mathbf{k_1}-\mathbf{k_2}, \,\, \,
 \mathbf{K}=(\mathbf{k_1}+\mathbf{k_2})/2\, \, .
\end{equation}
The factor `1/2' in Eq.~(\ref{def}) arises 
from the averaging over the
spins of the photons in the final state. 
 The space-time emission function $S$ is
approximated as the photon production rate,  
$EdN/d^4x d^3k$, from the quark and the hadronic 
matter phases. The source function $S$ is expressed as $S_Q+S_H$ in the 
numerator, where $Q$ and $H$ stand for quark and hadronic phases 
respectively. The contributions of the
quark matter or the hadronic matter or both to the correlation of
photons are estimated by retaining $S_Q$ or $S_H$ or both.
 
The correlation function 
$C({\bf q},{\bf K})$  is often represented
in terms of the outward, side-ward,  
and longitudinal momentum differences, $q_{o}$, 
$q_{s}$ and $q_{\ell}$. We can write the 4-momentum ($k_i^\mu$) 
of the ${\it{i}}$th photon as: 
\begin{equation}
k_i^\mu=(k_{iT}\, \cosh \, y_i, k_{iT}\,\cos \psi_i,k_{iT} \sin \psi_i,k_{iT} \, \sinh\, y_i)
\end{equation}
%
where $k_T$ is the transverse momentum, $y$ is the rapidity, 
and $\psi$ is the azimuthal angle. 
The difference of transverse momenta is defined as
\begin{equation}
\mathbf{q_T}=\mathbf{k_{1T}}-\mathbf{k_{2T}} \, \, ,
\end{equation}
and the average transverse momentum
\begin{equation}
\mathbf{K_T}=(\mathbf{k_{1T}}+\mathbf{k_{2T}})/2\, .
\end{equation}
In terms of these variables, the longitudinal,
outward, and side-ward momentum differences 
are obtained as~\cite{dks1}:
\begin{eqnarray}
q_{\ell}&=&k_{1z}-k_{2z}\nonumber\\
        &=&k_{1T} \sinh y_1 - k_{2T} \sinh y_2~,\\
\label{q_l}
q_{o}&=&\frac{\mathbf{q_T}\cdot \mathbf{K_T}}{K_T}\nonumber\\
       &=& \frac{(k_{1T}^2-k_{2T}^2)}
         {\sqrt{k_{1T}^2+k_{2T}^2+2 k_{1T} k_{2T} \cos (\psi_1-\psi_2)}}~,\\
\label{q_o}
q_{s}&=&\left|\mathbf{q_T}-q_{o}
       \frac{\mathbf{K_T}}{K_T}\right|\nonumber\\
        &=&\frac{2k_{1T}k_{2T}\sqrt{1-\cos^2(\psi_1-\psi_2)}}
      {\sqrt{k_{1T}^2+k_{2T}^2+2 k_{1T} k_{2T} \cos (\psi_1-\psi_2)}}~ .
\label{q_s}
\end{eqnarray}
The radii corresponding to these momentum differences are 
obtained by approximating the correlation function as,
\begin{eqnarray}
 C(q_{o}, q_{s},q_{\ell}) = 1 +  \frac{1}{2}\exp 
\left[-\left(q_{o}^2R_{o}^2 +  q_{s}^2R_{s}^2 
+ q_{\ell}^2R_{\ell}^2 \right ) \right].
\end{eqnarray}
The root mean square momentum difference $\langle q_i^2 \rangle$ is,
\begin{eqnarray}
\langle q_i^2 \rangle \, = \, \frac {{\displaystyle \int} \, (C-1)\, q_i^2 \, dq_i} { {\displaystyle \int} \, (C-1) \, dq_i},
\label{qi}
\end{eqnarray}
and for the Gaussian parameterization shown above, this becomes
\begin{eqnarray}
R_i^2 \, = \, \frac{1}{2 \, \langle q_i^2 \rangle} \, .
\end{eqnarray}

\begin{figure}[tb]
\includegraphics[ width=7.0cm, clip=true]{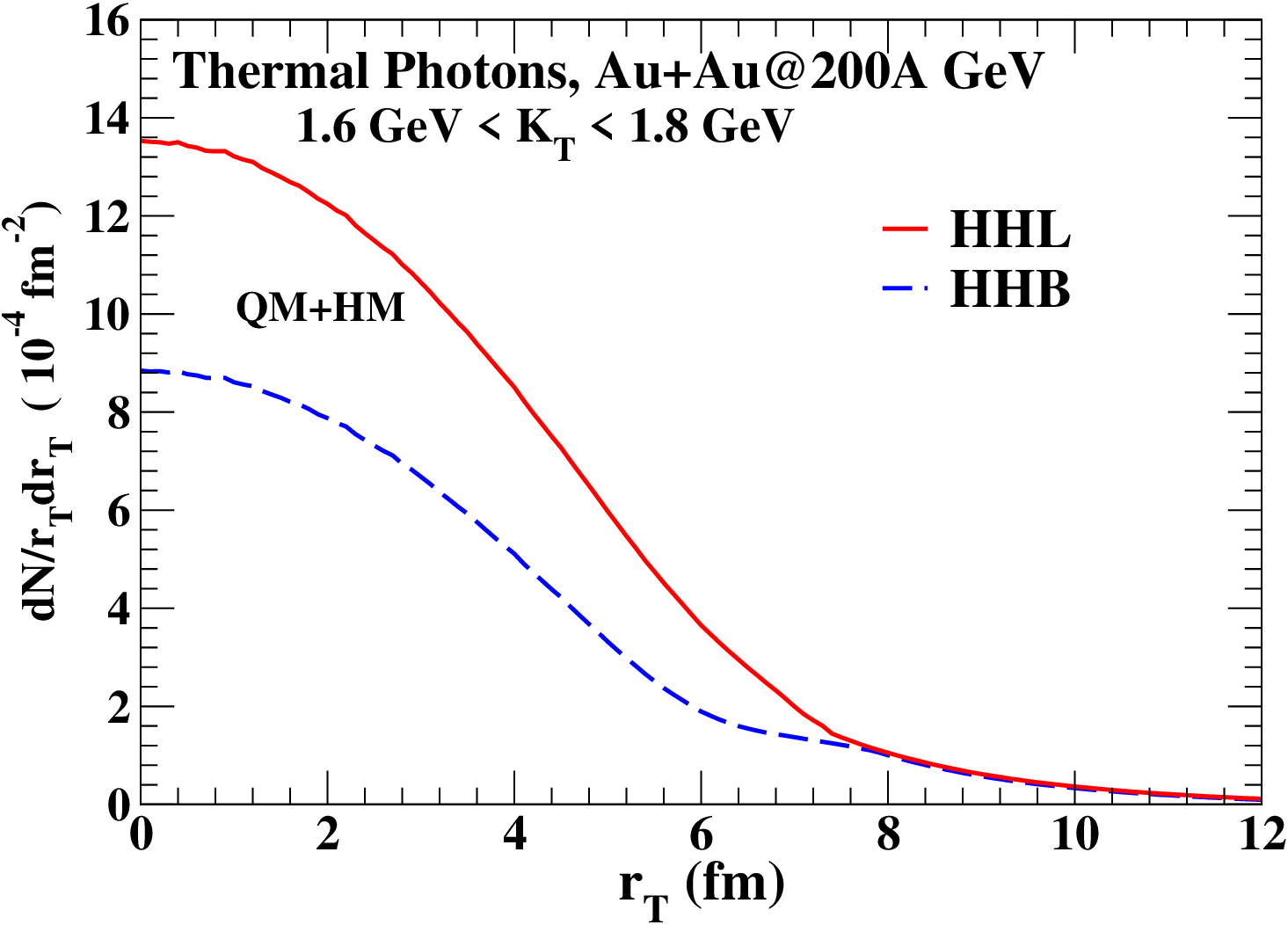}
\includegraphics[ width=7.0cm, clip=true]{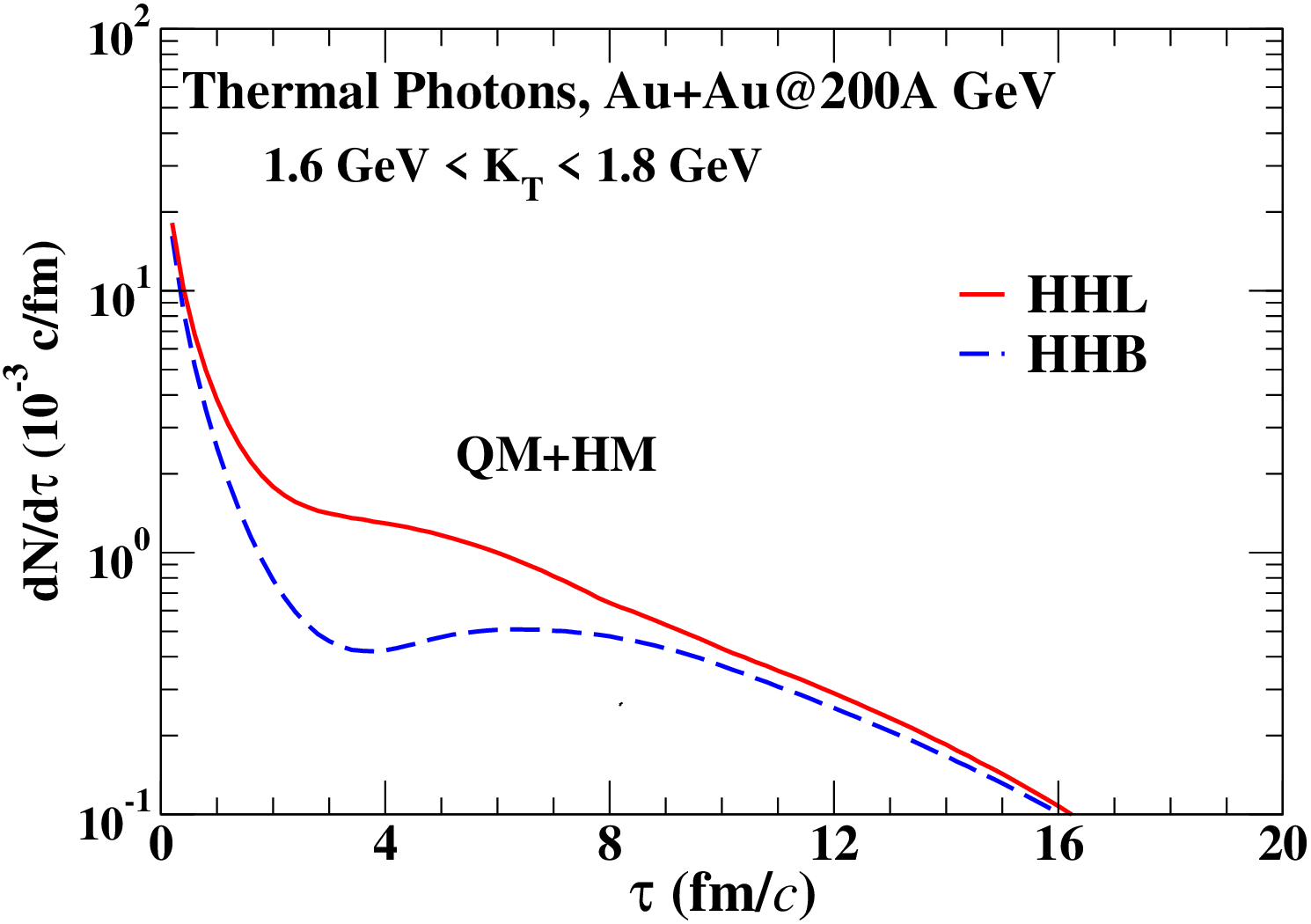}
\caption{(Colour on-line) The radial (left panel) and temporal dependence
(right panel)
of the source of photons for HHL and HHB equations of state for central
collision of gold nuclei at the top RHIC energy.}
\label{fig7}
\end{figure}

As a first step we plot the spatial and the temporal distributions of
the sources obtained for the two equations of state in Fig.~\ref{fig7},
for a typical momentum of the photons $K_T \approx$ 1.7 GeV,
which provide valuable information about the dynamics of the system.

We find that the HHL equation of state leads to a larger
production of photons at smaller radial distances 
as well at intermediate times. The spectrum of the photons in Fig.~\ref{fig4},
already indicates this. The photons from
quark matter originate at early times and those from hadronic
matter at later times, and as the interference between
the two depends on their relative contributions~\cite{dks3}, this difference
holds out a hope that we may see a difference in the intensity
interferometry  of photons for the two equations of state.

In order to bring out the dependence of the correlation function on the outward,
sideward, and longitudinal momentum differences very clearly, we choose momenta
of the interfering photons such that when $q_o \ne 0$, then  $q_s$ and $q_l$ are
zero, and so on. 
 
\begin{figure}[tb]
\begin{center}
\includegraphics[width=7.5cm, clip=true]{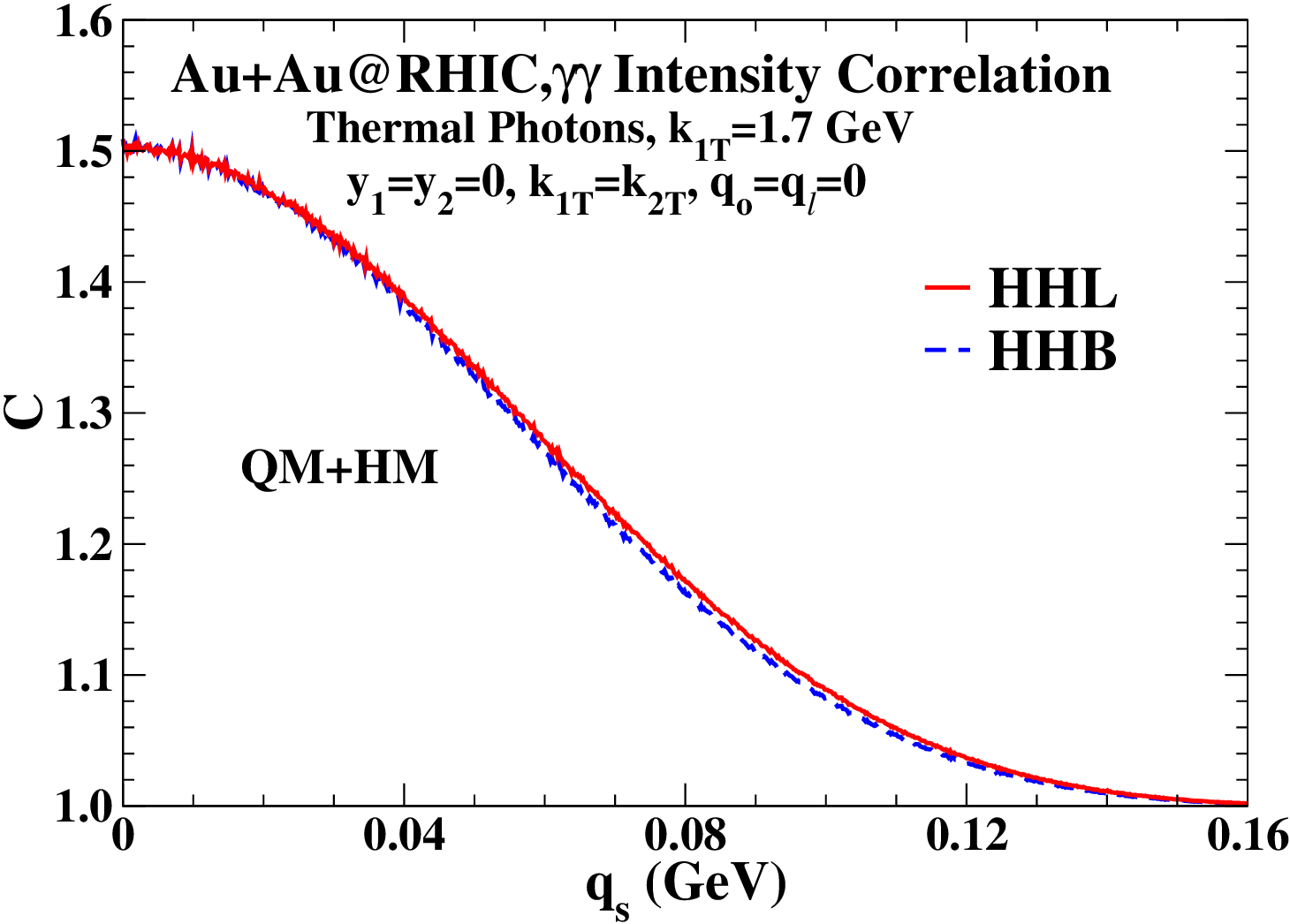}
\includegraphics[width=7.5cm, clip=true]{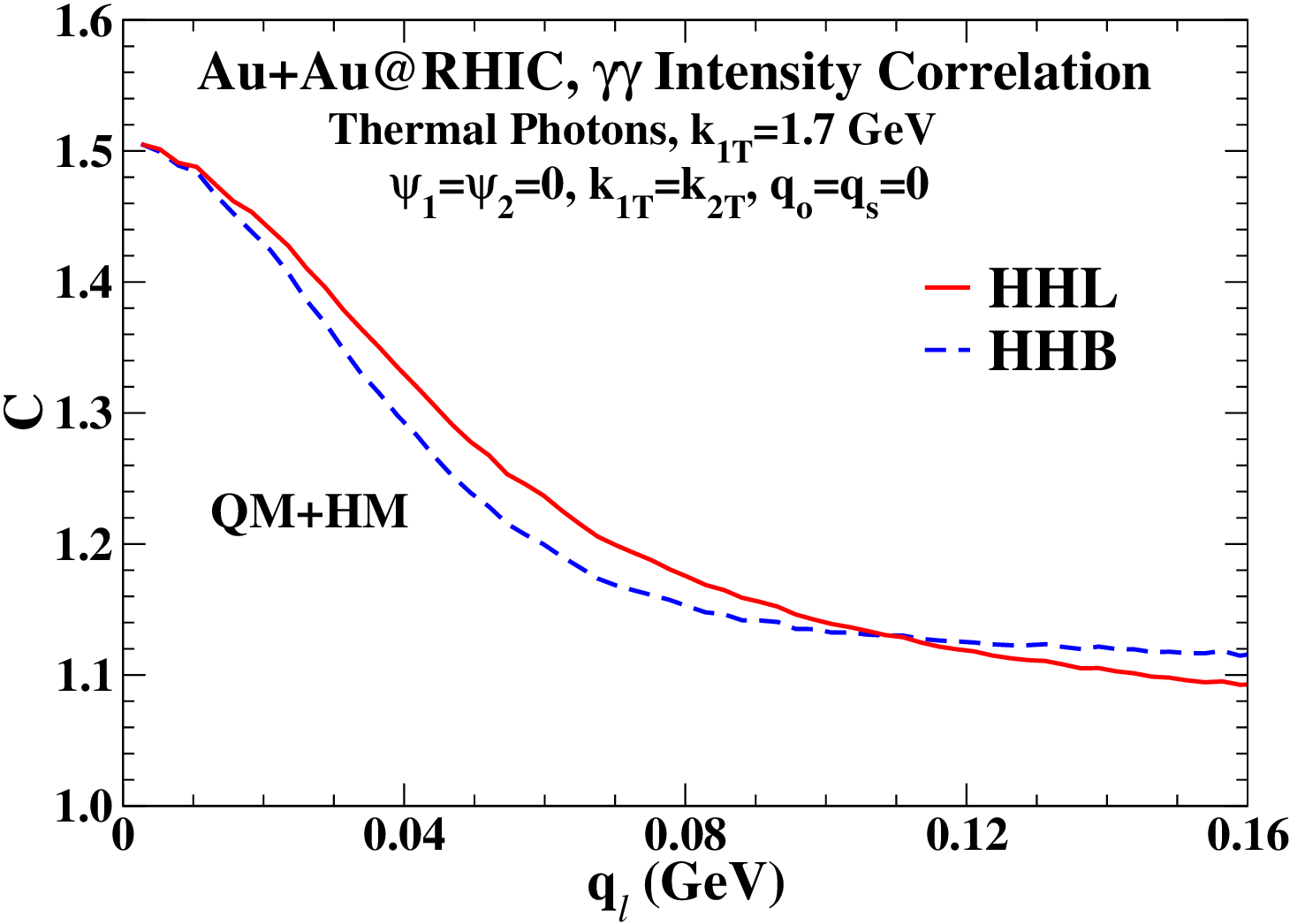}
\end{center}
\caption{(Colour on-line) The sideward (left panel)
and longitudinal correlation (right panel)
functions for thermal photons produced in central collision
of gold  nuclei at RHIC.}
\label{fig8}
\end{figure}

The corresponding side-ward and longitudinal correlation functions are shown 
in Fig.~\ref{fig8}. We see that while the sideward correlation is essentially
identical for the two equations of state, the longitudinal  
correlation function shows a mild variation when we change the
 equation of state.

\begin{figure}[t]
\begin{center}
\includegraphics[width=7.5cm, clip=true]{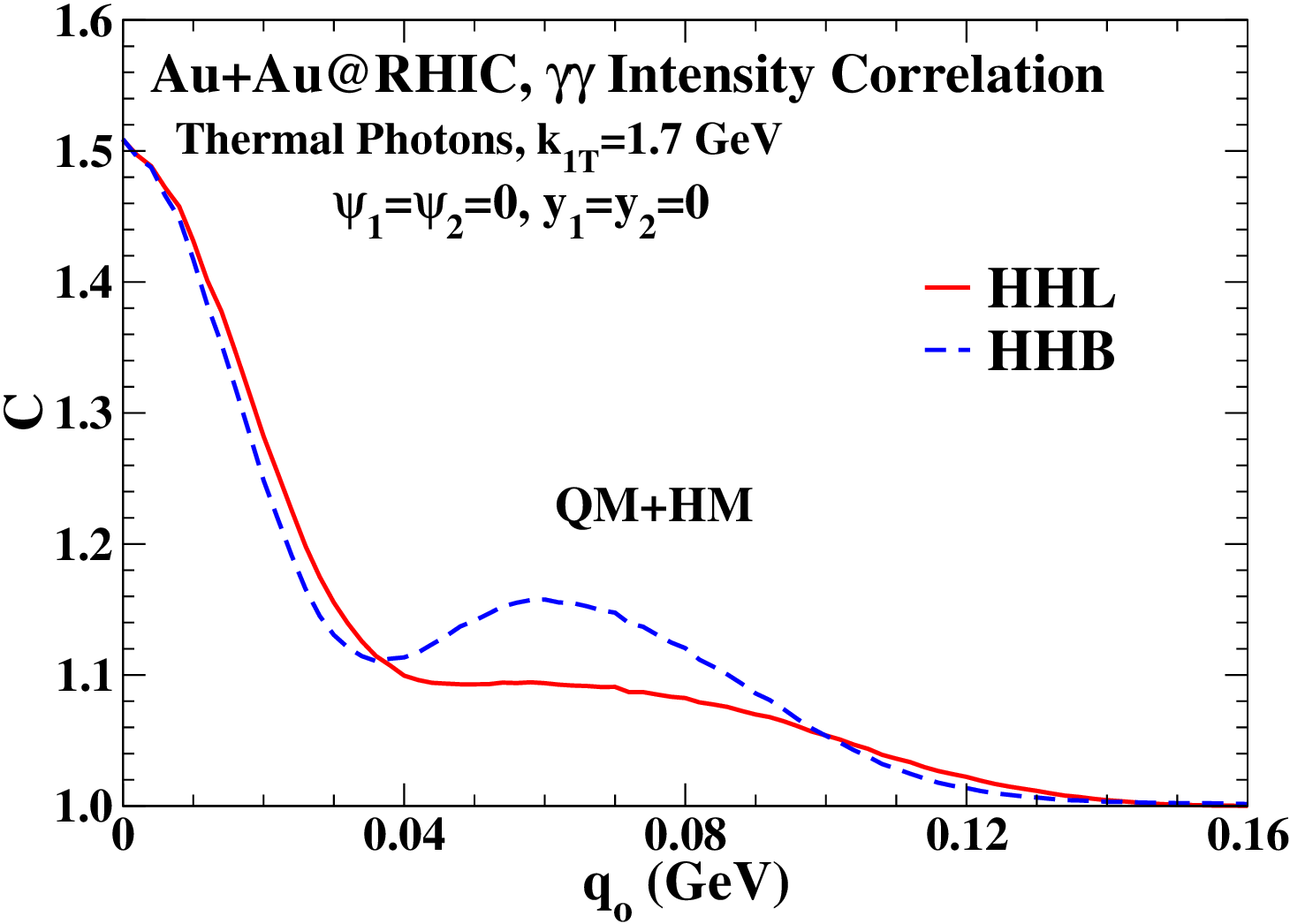}
\includegraphics[width=7.5cm, clip=true]{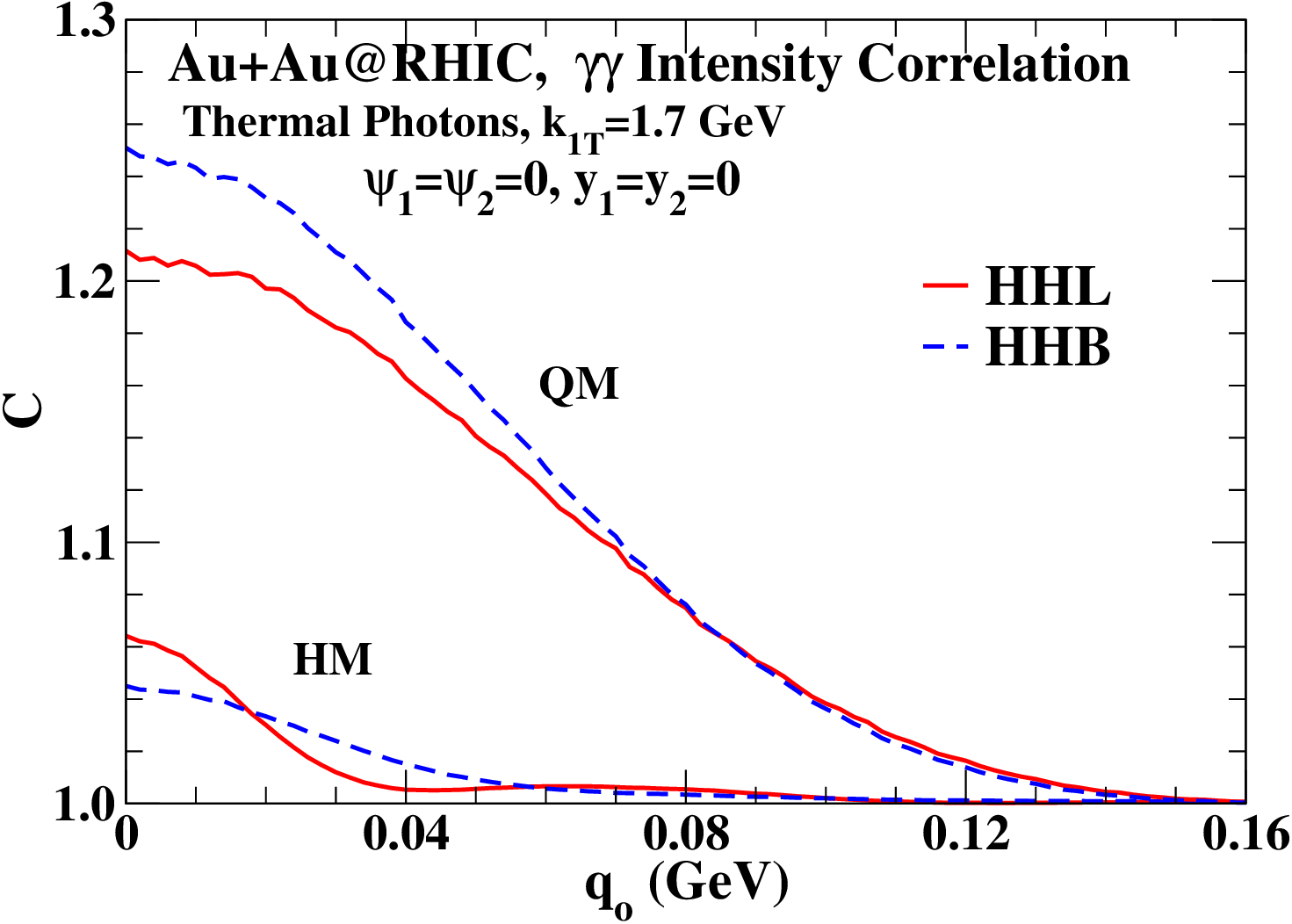}
\end{center}
\caption{(Colour on-line) Left panel: The outward correlation
function for thermal photons produced in central collision
of gold nuclei at RHIC. Right panel: Individual contributions of quark
matter and hadronic matter to the correlation.}
\label{fig9}
\end{figure}

The results for the outward correlation (Fig.~\ref{fig9})
are more dramatic and along with the interference already reported
in Ref.~\cite{dks3}, we see a very clear difference in the
results obtained for the equations of state. Recalling that this
interference is decided by the relative contribution of the hadronic
and quark phase, we also show the quark matter and hadronic matter 
contributions to the correlation function. We notice a clear difference in the
underlying radii for the hadronic matter contribution for the 
equations of state; the
one for the HHB equation of state being effectively much larger.

 We note that in general the experimentally measured correlation
functions (see, e.g., Ref.~\cite{wa98}) have larger error-bars for smaller
values of the momentum differences. However, these  measurement
normally have smaller error-bars
for larger values of momentum differences, where our results are
seen to be most sensitive to the differences in the equations of 
state. 
This holds out the hope that these differences can be experimentally
verified.

We follow the treatment of Ref.~\cite{dks3} to 
estimate the source-sizes of the HM and QM contributions along with
their separation (see later) and write:
\begin{eqnarray}
C(q_i, \alpha) = 1 \ + \ 0.5 |\rho_{i,\alpha}|^2 
\end{eqnarray}
where $i$ = $o$, $s$, or $\ell$, and $\alpha$ denotes quark matter (Q) 
and hadronic matter (H) in an obvious notation. 
 
Next we approximate the source distributions $|\rho_{i,\alpha}|$ as a 
Gaussian and write:
\begin{eqnarray}
 |\rho_{i,\alpha}| \ = \ I_\alpha \ \exp \left [- \ 0.5 \ 
(q_i^2R_{i,\alpha}^2) \right ] 
\end{eqnarray}
where $I_Q = dN_Q/(dN_Q+dN_H)$ and  
$ I_H = dN_H/(dN_Q +dN_H)$. The
final correlation functions 
are then approximated as,
\begin{eqnarray}
C(q_i) \ = \  1 \ & + & \ 0.5 \left[ \ |\rho_{i,Q}|^2 \ +  \ 
|\rho_{i,H}|^2 \  +  \ 2 \ |\rho_{i,Q}| |\rho_{i,H}| \ \cos (q_i\Delta R_i) 
\right] 
\label{cc} 
\end{eqnarray} 
which clearly reveals the interference between the two 
sources. Here $\Delta R_i$ stands for the 
separation of the two sources.
 Thus for example, for thermal photons having $K_T 
\approx 1.7$ GeV at RHIC, the radii (in fm) corresponding to each source 
are obtained as: 

\vskip 0.2cm 

\textbf{HHB}
\begin{equation}
R_{o,Q} =  2.5, \ R_{o,H}  =  8.3, 
 \ \Delta R_o  = 14.9~. 
\label{ex1}
\end{equation}

\vskip 0.2cm 



\textbf{HHL}
\begin{equation}
R_{o,Q} =  2.7, \ R_{o,H}  =  4.8,  
\ \Delta R_o  = 13.4~.
\label{ex2}
\end{equation}

\begin{figure}[t]
\includegraphics[ width=7.0cm, clip=true]{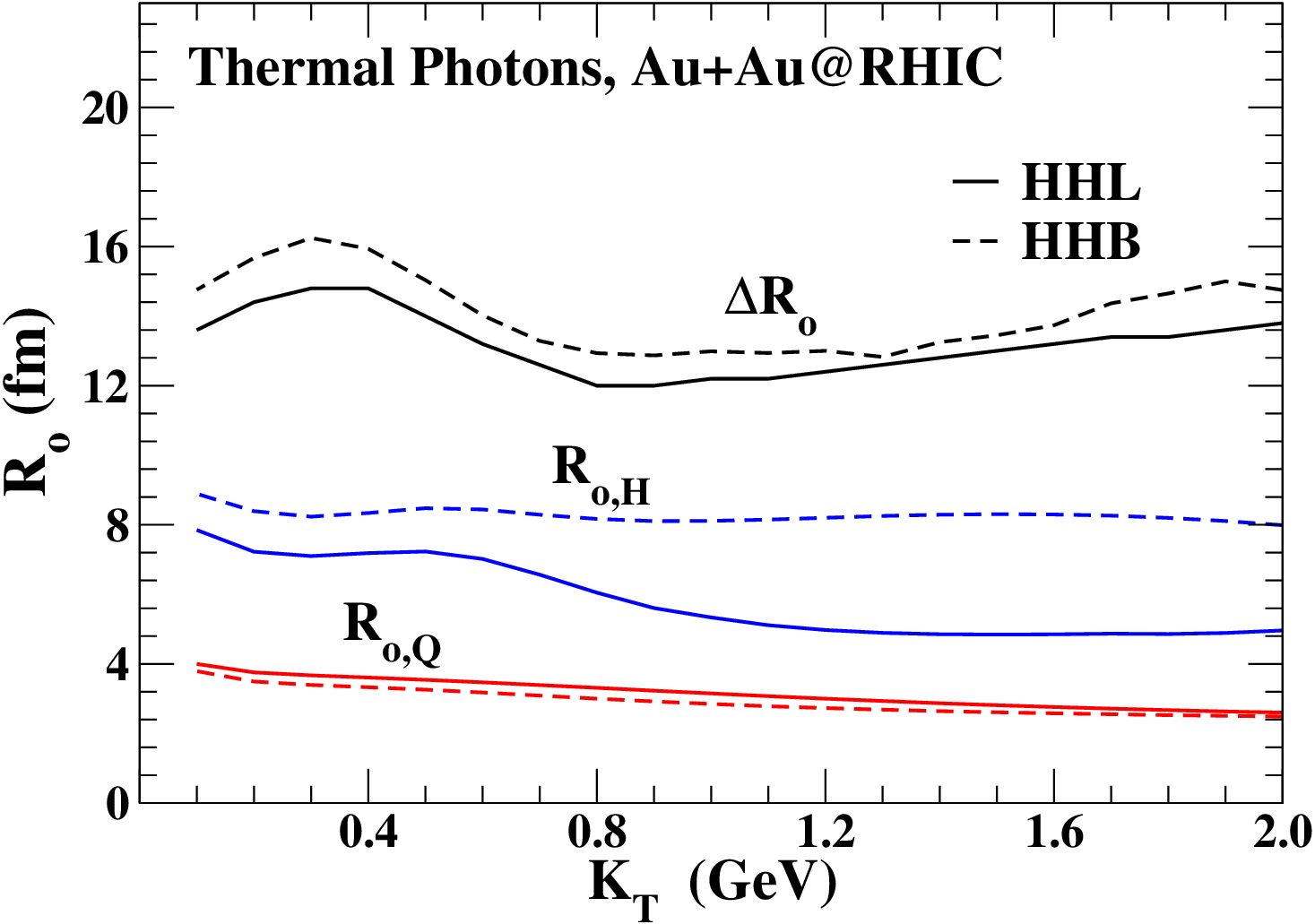}
\includegraphics[ width=7.0cm, clip=true]{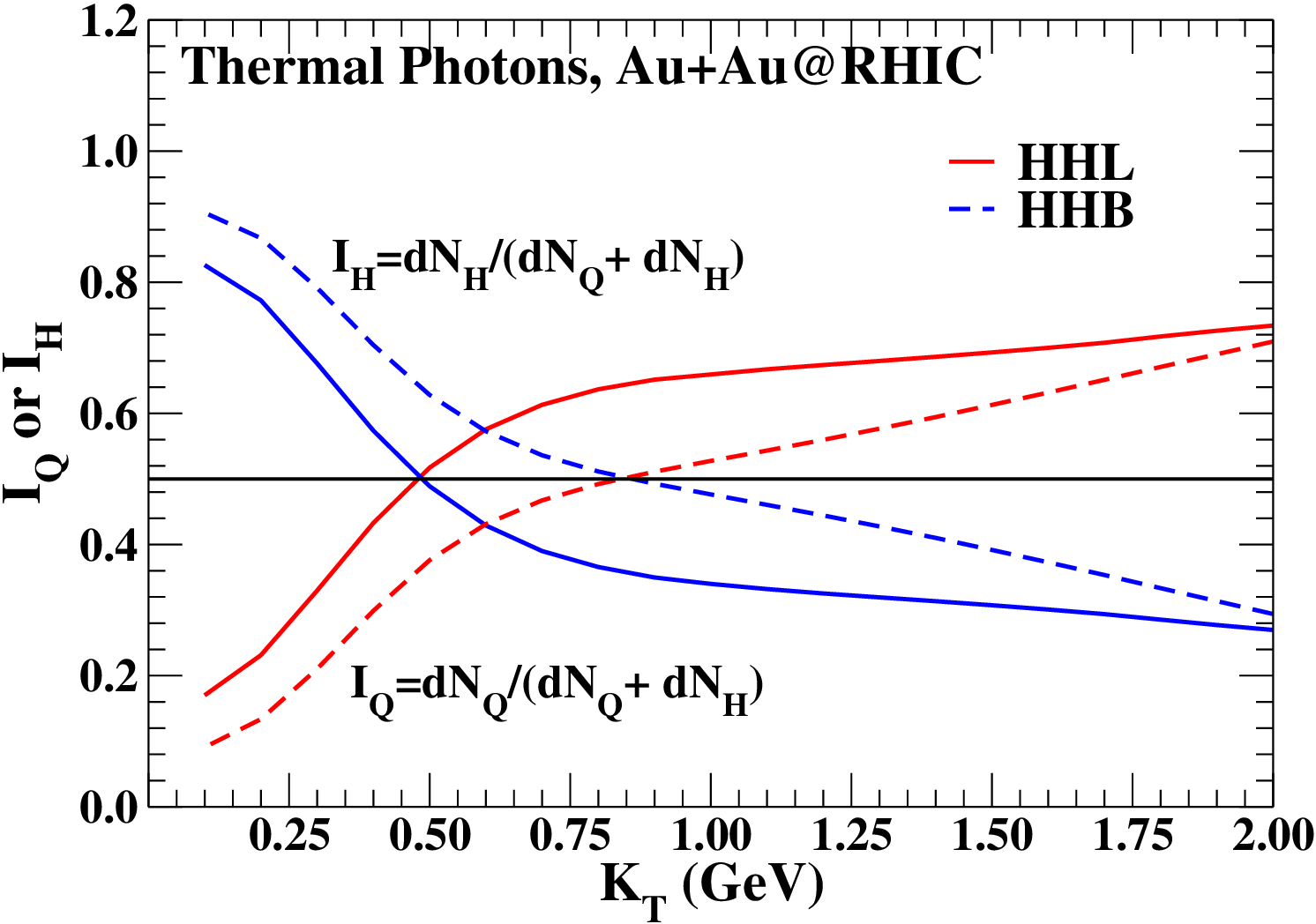}
\caption{(Colour on-line) Left panel: Transverse momentum dependence 
of outward radii for  hadronic and quark matter sources 
for thermal photons. Right panel: The 
fraction of thermal photons emitted from quark matter and
hadronic matter. }
\label{fig10}
\end{figure}

\vskip 0.2cm 
 The transverse momentum dependence of the
radii and the separation $\Delta R$ for the outward
correlation function is shown in the left panel
of Fig.~\ref{fig10}.
We see that the separation $\Delta R_o$, which corresponds
to the life time of the system is slightly smaller for the HHL
equation, as it does not incorporate a mixed phase (see discussion in
Ref.~\cite{dks3}).
One can obtain similar results for the contributions
for the sideward and longitudinal radii. The $R_o$ for the hadronic matter
for the HHL equation of state is also much smaller for the same reason.    

The transverse momentum dependence of the quark and the 
hadronic matter fractions $I_Q$ and $I_H$ are shown in the right
panel of Fig.~\ref{fig10} 
where, $I_Q$ for HHL is found to be larger than that for HHB.
As the speed of sound for the HHL equation of state is smaller
during the QGP phase (see Fig.~\ref{fig3}), the system cools
 at a slower rate,  which leads to a larger contribution from the quark phase.
We also find that for HHL equation of state, the quark and hadronic matter contributions become equal ($I_Q=I_H=0.5$) at a slightly lower $K_T$.

Next we consider correlation of thermal
photons from central collision of lead nuclei at LHC. 
\begin{figure}[tb]
\begin{center}
\includegraphics[width=7.5cm, clip=true]{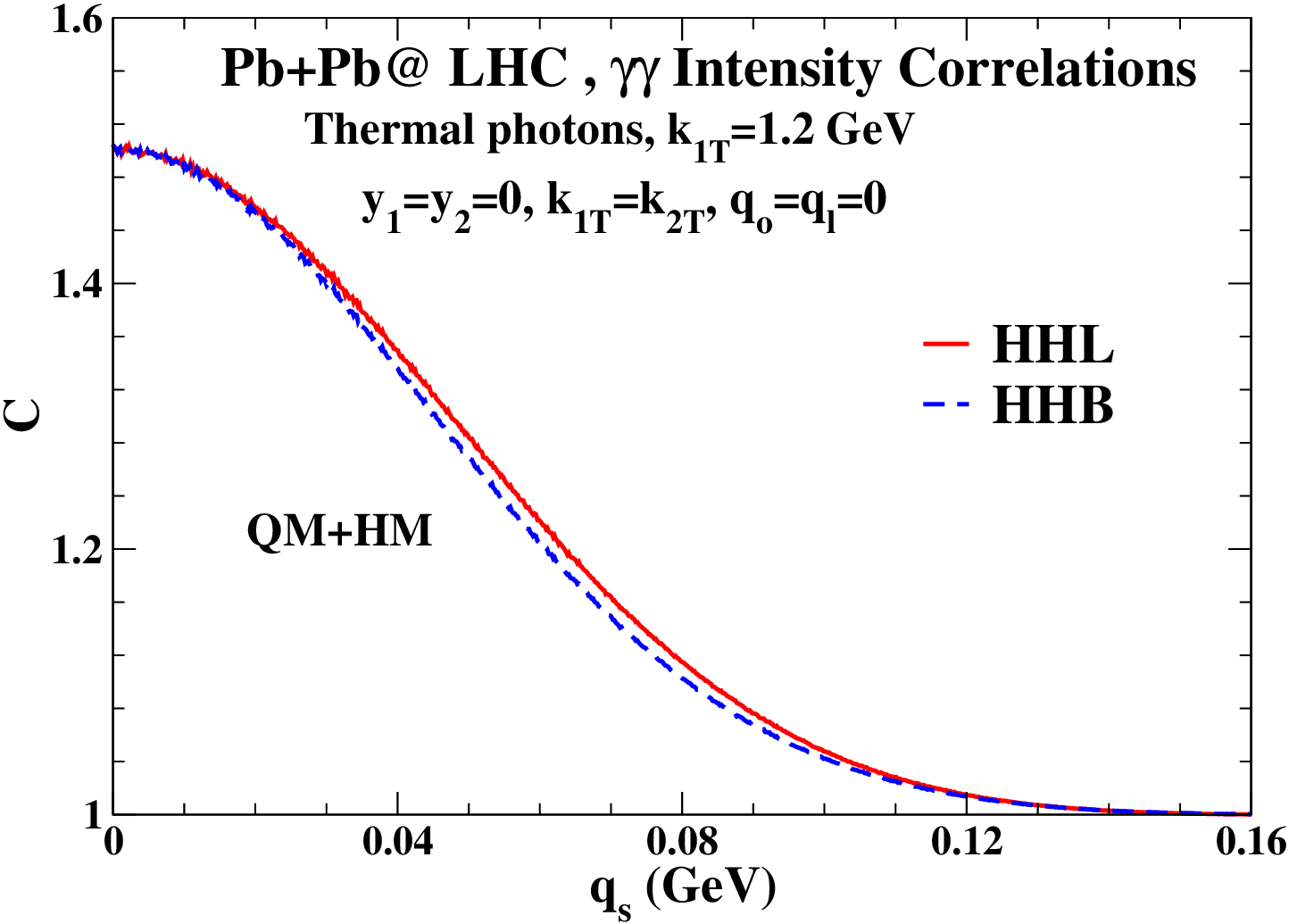}
\includegraphics[width=7.5cm, clip=true]{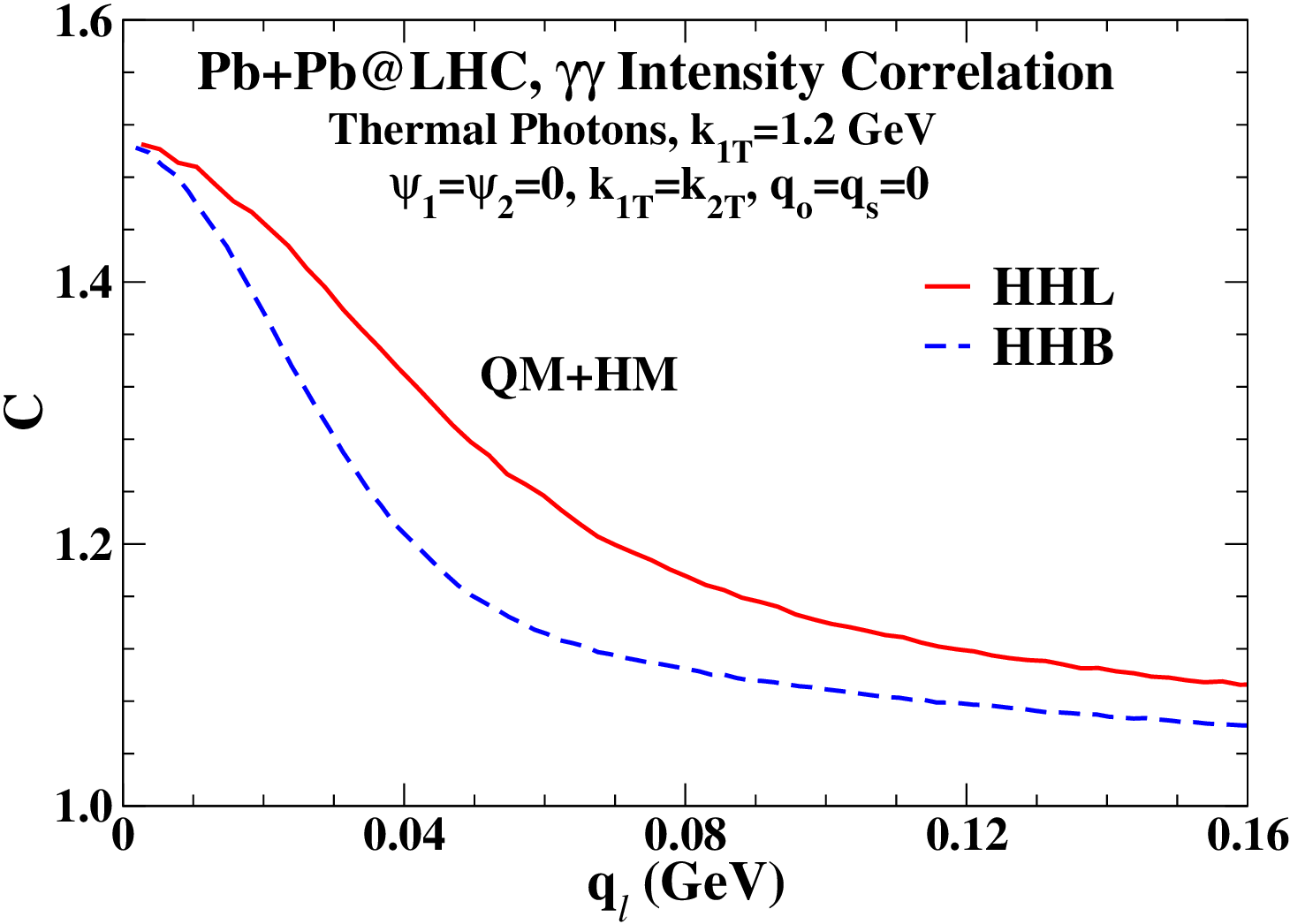}
\end{center}
\caption{(Colour on-line) The sideward (left panel)
and longitudinal correlation (right panel)
functions for thermal photons produced in central collision
of lead  nuclei at LHC.}
\label{fig11}
\end{figure}

The  results for the sideward and the longitudinal correlations for 
transverse momentum of about 1.2 GeV are shown in Fig.~\ref{fig11}.
It is observed that, while the sideward correlation is
again nearly
identical for the two equations of state, the longitudinal correlation
is fairly different and quite sensitive to the equation of state
used. In view of the importance of this observation it
 would be interesting to analyze this behaviour
using a 3+1D hydrodynamics~\cite{chiho}.

\begin{figure}[tb]
\begin{center}
\includegraphics[width=7.5cm, clip=true]{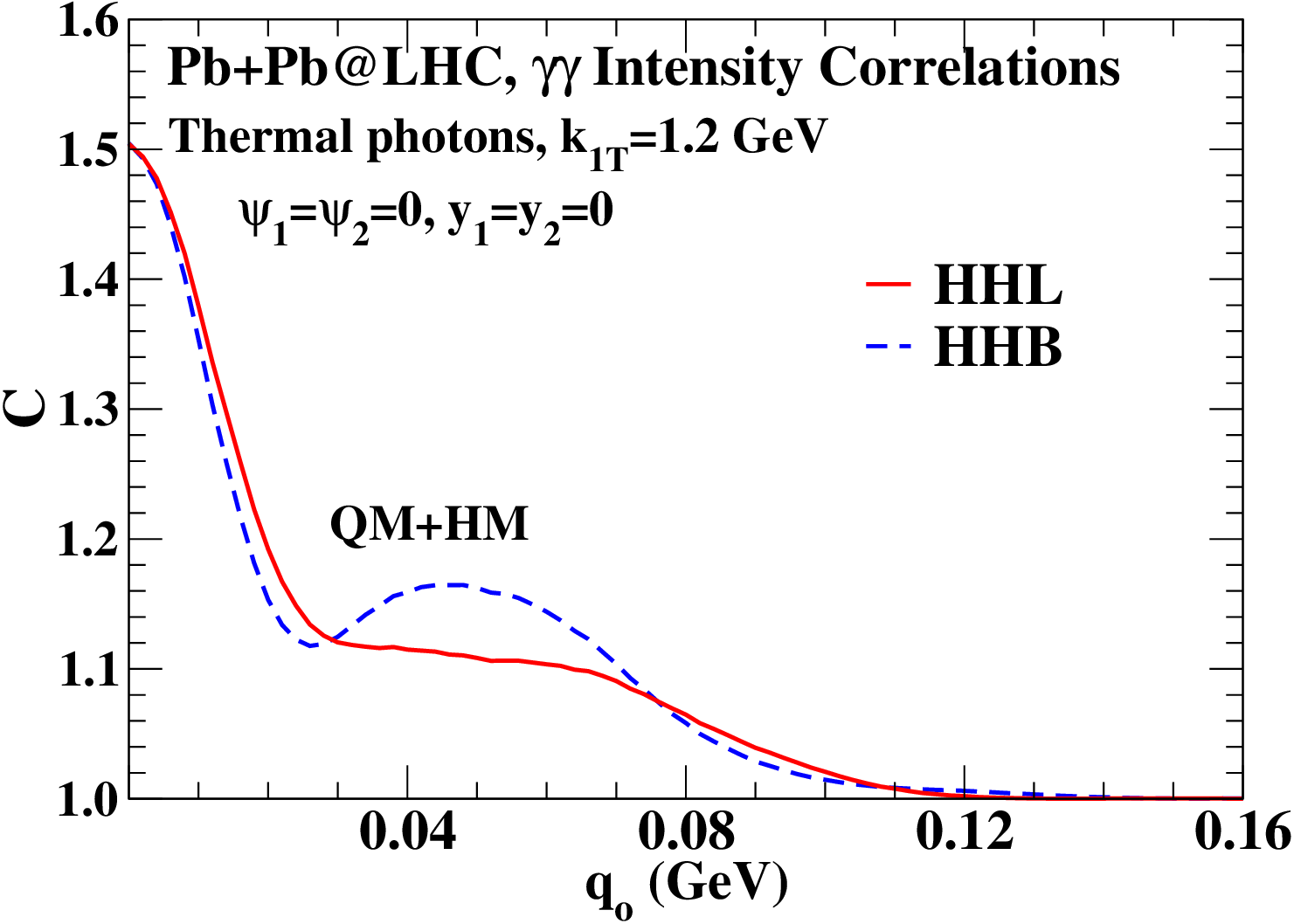}
\includegraphics[width=7.5cm, clip=true]{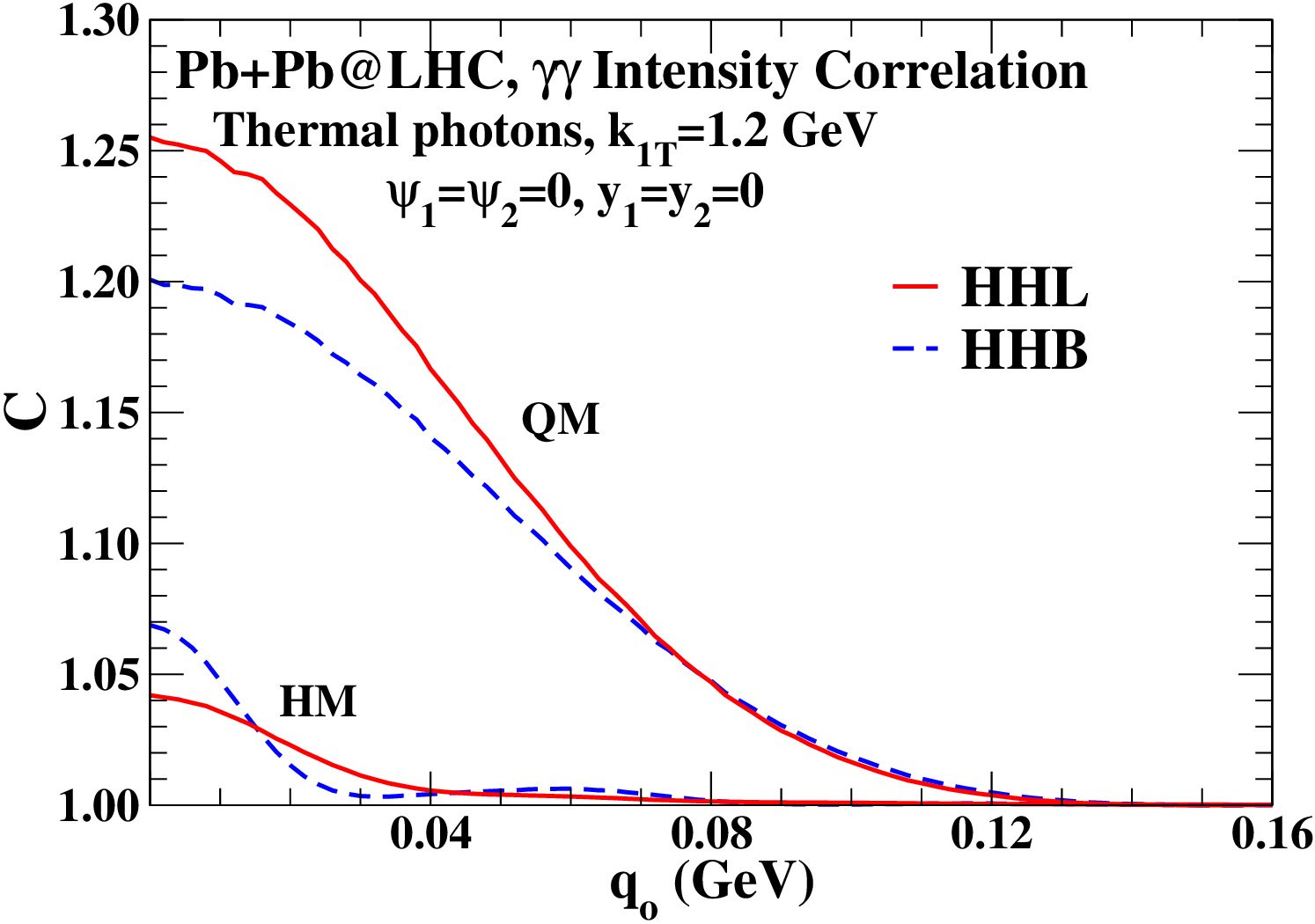}
\end{center}
\caption{(Colour on-line) The outward correlation
functions for thermal photons produced in central collision
of lead nuclei at LHC for equation of state HHB and HHL. }
\label{fig12}
\end{figure}

The results for the outward 
correlation function and its composition in terms
of contributions from the quark matter and the hadronic matter are
shown in Fig.~\ref{fig12}. Once again 
a very distinctive sensitivity is observed for the two equations of state.

\section{Summary and Discussions}

In brief, we have constructed a rich description of hadronic matter
at zero baryonic chemical potential by including all the mesons
having $M <$ 1.5 GeV, and all the baryons having $M<$ 2 GeV,
in thermal and chemical equilibrium along with Hagedorn  resonances.
Incorporation of finite volume for the hadrons
gives results, which join smoothly with those with the recent
lattice calculations. We construct two equations of state by
switching over to a bag model equation of state (leading to
a first order phase transition) or to the lattice
equation of state (showing a rapid crossover around 185 MeV)
 at a temperature of 165 MeV. 

We find that spectra of pions, kaons, protons and thermal photons 
both at RHIC and LHC energies are quite similar for a given 
initial condition. However, the experimental data at RHIC 
show a  slight preference for the equation of state incorporating 
lattice results. 

Still, the longitudinal correlation to a lesser extent and the
outward correlation to a great extent, for thermal photons
are found to be sensitive to the equation of state.
 If verified in experiments,
this could prove to be a very valuable probe for the equation of
state of strongly interacting matter. We have explained these
observations by having a close look at the history of evolution of
the systems.

Before closing, we note that  these studies could
be improved in some ways. One could, for example, investigate the consequences of inclusion
of viscous effects and using a 3+1D dimensional hydrodynamics,
especially for the longitudinal correlation. This will
also free our findings from the constraint of the longitudinal
 boost-invariance assumed here. It would also be of interest to extend this
study to non-central collisions, which we shall do shortly. 
Consequences of varying some of the initial conditions, e.g., formation
time and colour gluon condensate inspired initial conditions could
 also be studied. It may  also be worth-while to vary the transition 
temperature for the two equations of state and study its consequences. 

Future directions would involve extending these studies to non-zero
baryonic chemical potential, especially in view of the low energy runs at
RHIC and upcoming studies at FAIR, though we await more accurate
lattice results for non-zero $\mu_B$. 

In a forthcoming publication we shall extend the present work to study the
 elliptic flow of dileptons as well as the elliptic flow of photons~\cite{rupa2}.  

\section*{Acknowledgements} Two of us, SD and RC are grateful to
Department of Atomic Energy for financial support during the course of
this work. We thank M. Cheng, S. Dutta, and S. Gupta
 for useful correspondence.

\end{document}